\documentclass{IEEEtran}

\usepackage{cite}
\usepackage{amsmath,amssymb,amsfonts}
\usepackage{algorithmic}
\usepackage{graphicx}
\usepackage{textcomp}
\usepackage{xcolor}
\usepackage{url}
\usepackage[normalem]{ulem}
\usepackage[export]{adjustbox}
\usepackage{stfloats}
\usepackage{lipsum}
\usepackage{subfigure}
\usepackage{cuted}
\usepackage{tabularx}
\usepackage{mathtools}
\usepackage{enumitem} 

\def\BibTeX{{\rm B\kern-.05em{\sc i\kern-.025em b}\kern-.08em
    T\kern-.1667em\lower.7ex\hbox{E}\kern-.125emX}}

\begin{document}
\title{Metagratings on Low-Cost Substrates for Efficient Anomalous Reflection: Addressing Dielectric Loss}
\author{Oz Diker and Ariel Epstein, \IEEEmembership{Senior Member, IEEE}
\thanks{Manuscript submitted on January 1st, 2026.}
\thanks{The authors are with the Andrew and Erna Viterbi Faculty of Electrical
and Computer Engineering, Technion–Israel Institute of Technology,
Haifa, 3200003, Israel (email: oz.diker@campus.technion.ac.il; epsteina@
ee.technion.ac.il)}}
\maketitle

\begin{abstract}
We present a theoretical framework and practical methodology for designing high-efficiency metagratings (MGs), sparse periodic arrangements of subwavelength polarizable particles (meta-atoms), on low-cost dielectric substrates with non-negligible losses. The formulation incorporates these losses and exploits multiple degrees of freedom to optimize beam manipulation efficiency within a simple realistic printed-circuit-board (PCB) configuration. Importantly, the various loss mechanisms are analyzed using a judiciously devised equivalent circuit model, providing insights on their respective contributions. 
We validate our theory by designing, fabricating, and experimentally characterizing an efficient FR4-based anomalous reflection PCB MG, demonstrating good agreement between analytical predictions, full-wave simulations, and laboratory measurements. This work opens avenues for realizing efficient, low-profile, beam manipulation devices at reduced cost, offering practical solutions to mitigate loss limitations in diverse material sets across the electromagnetic spectrum.
\end{abstract}

\begin{IEEEkeywords}
Metagratings, anomalous reflection, low-cost substrate, equivalent circuit model, dielectric loss, FR4
\end{IEEEkeywords}

\section{Introduction}
\label{sec:introduction}
Recent years have seen significant progress in the field of beam manipulation by compact complex media platforms, with particular advancements in metagratings (MGs) \cite{Radi2021}. MGs are periodic devices characterized by a sparse arrangement of subwavelength polarizable particles (meta-atoms), intended for intricate diffraction engineering of the incident field. 
The wavefront scattered off the MG can be described as a superposition of discrete spatial harmonics, in accordance with Floquet-Bloch (FB) theory. Through meticulous design of the particle locations, shapes, and dimensions, the scattering pattern can be precisely controlled, by determining the extent of coupling to the various FB harmonics {\cite{ra2017metagratings,sell2017large}. 
This control, implemented by the collective response of the meta-atoms, has paved the way for innovative applications in optics, microwaves, and terahertz technologies, facilitating demonstrations of MGs for highly efficient wave manipulation functionalities, such as anomalous reflection and refraction, beam steering, and focusing, to name a few 
\cite{ra2017metagratings,memarian2017wide,sell2017large,wong2018perfect,epstein2017unveiling,PaniaguaDominguez2018,Kang2020,popov2021non,yang2023terahertz}.

Specifically, loaded-wire MGs fabricated using printed circuit board (PCB) technology have gained significant attention for device realization at microwave frequencies \cite{epstein2017unveiling, Rabinovich2018, Epstein2018, Popov2018, 8892735}. These PCB-based MGs utilize patterned elongated metallic strips as meta-atoms, arranged on a dielectric substrate or embedded within a stack of such planar dielectric layers (possibly backed by a metallic ground plane if reflection mode operation is desired). 
In such a configuration, the patterned metal traces (meta-atoms) may be modelled analytically as flat wires uniformly loaded by effective impedance per unit length, the value of which is governed by the specific geometric properties of the load pattern \cite{tretyakov2003analytical, Ikonen2007}. 
Correspondingly, to operate as effective beamformers, such MGs are designed as to develop prescribed induced currents within the conductors in response to the incident field excitation, established via these analytical models \cite{epstein2017unveiling, 8892735}. These, in turn, allow for the tailored manipulation of the scattering pattern of the reflected or transmitted field, enabling advanced wavefront engineering capabilities, e.g., for beam deflection, guided mode manipulation, and dynamic antenna devices \cite{Popov2018, xu2020dual, Popov2020_1, killamsetty2021metagratings, biniashvili2021eliminating, popov2021non, xu2022extreme, 9957042, Hu2023, li2024reconfigurable, 10932694}. 

To date, studies have demonstrated the efficacy of PCB-based MGs mainly employing low-loss dielectric substrates in the microwave domain, yielding components with very high efficiencies \cite{8892735, Popov2020}. However, relying on such specialized substrates may be costly, and may even pose manufacturing hurdles due to limited availability. Consequently, to advance MGs as a primary platform for beam manipulation applications, it is essential to explore the feasibility of utilizing economical materials for design and implementation, while ensuring minimal impact on the device performance \cite{9415622}. More affordable materials often exhibit higher losses (e.g., FR4 \cite{667012}) compared to those typically used in previously reported high-efficiency prototypes ($\tan\delta$ in order of 0.001). Hence, to assess the viability of using cost-effective materials, it is necessary to investigate the impact of substrate losses, manifested by the imaginary part of the laminate's complex permittivity, and evaluate the related performance degradation of the MG. Such an analysis will provide insights into the trade-offs between cost and efficiency, guiding the development of more economical -- yet effective -- PCB-based MGs, while highlighting the relevant physical processes and their potential impact.

In this paper, we fill this gap, presenting a comprehensive analysis of the loss mechanisms in such low-cost PCB-based MGs and develop a corresponding design methodology to mitigate them. To achieve the latter, we emphasize the necessity of incorporating additional degrees of freedom into the design, implemented by integrating an additional meta-atom within the MG period \cite{diker2023low}.
Importantly, we propose an equivalent circuit model that accounts for both dielectric and conductor losses and their impact on the load impedance, providing crucial physical insight regarding the different contributions to power dissipation, \textcolor{black}{leading to improved} design guidelines. 
We validate our analysis and synthesis through a case study focused on the realization of an FR4-based MG designed for anomalous reflection, achieving efficiencies greater than $85\%$ in the K-band, despite the non-negligible losses ($\tan\delta = 0.02$). Our analytical findings are validated through full-wave simulations and experimental measurements, showcasing the effectiveness of our approach in maintaining the trademark high-efficiency diffraction engineering capabilities of PCB-based MGs, even when relying on low-cost substrates. 
These results not only shed light on the effects of the various loss processes on MG modeling and performance, but also provide practical solutions to optimize their operation in real-world applications. Equally important, these revelations may enable intentional harnessing of power dissipation in the substrate to further enhance the performance of recently proposed MG absorbers \cite{yashno2022large,boust2022metagrating,10012668}.
\begin{figure*}[htb]
    \includegraphics[width=0.95\textwidth]{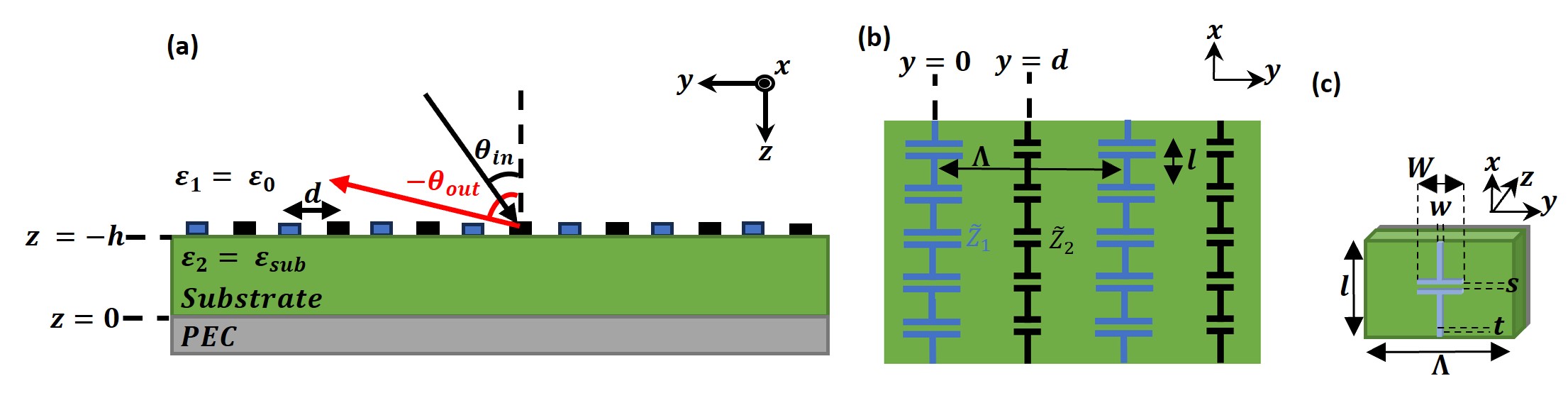}
    \caption{MG physical configuration. (a) Two different grids of loaded strips lie on a dielectric substrate backed by PEC and shifted by distance $d$ from each other. The MG excited by a TE polarized plane wave with an angle of incidence of $\theta_\mathrm{in}$.  
    (b) Top view of the MG, consisting of two different load impedance per unit length $\tilde{Z}_{1}$ and $\tilde{Z}_{2}$ per period, featuring periodicities $\Lambda$ and $l$ along $y$ and $x$, respectively. 
    (c) Zoom in on a single meta-atom, featuring a printed capacitor of width $W$ with copper traces of dimensions $s$ and $w$ and thickness $t$.}
    \label{fig:configuration}
\end{figure*}

\section{Theory}
\label{sec:theory}
\subsection{Scattered fields}
\label{subsec:scattered_fields}
We consider a 2D configuration ($\partial/\partial x = 0$), comprised of loaded copper strips of thickness $t$ and width $w$, serving as meta-atoms in our composite MG (Fig. \ref{fig:configuration}). These strips are arranged periodically on the plane $z=-h$, featuring a period length $\Lambda$ along the $y$ axis.
Each strip is loaded by printed capacitors repeating with periodicity $l\ll\lambda$ 
along the $x$ axis, effectively forming uniform load impedance density per unit length $\tilde{Z}_q$ (associated with the $q$th meta-atom). 
These metallic patterns are defined on a dielectric substrate of height $h$ (medium 2), characterized by permittivity $\varepsilon_{2} = \varepsilon_\mathrm{sub} = (\varepsilon'_\mathrm{2,r} - j\varepsilon''_\mathrm{2,r})\varepsilon_0$ with a perfect electric conductor (PEC) attached to its rear at the $z=0$ plane [Fig. \ref{fig:configuration}(a)]; the imaginary part, $\varepsilon''_\mathrm{2,r}$, is associated with substrate loss, and is usually described by the corresponding loss tangent, $\tan\delta = \varepsilon''_\mathrm{2,r}/\varepsilon'_\mathrm{2,r}$. The surrounding medium $z<-h$ (medium 1), has permittivity $\varepsilon_{1}$, assumed to be identical to the vacuum permittivity $\varepsilon_0$. For the $i$th 
medium, we define the wave number as $k_{i}=\omega\sqrt{\mu_{i}\varepsilon_{i}}$, and the wave impedance as $\zeta_{i}=\sqrt{\mu_{i}/\varepsilon_{i}}$. The described periodic structure, the MG, is excited by a transverse electric (TE) polarized plane wave ($E_{z}=E_{y}=H_{x}=0$) impinging at an angle of incidence $\theta_\mathrm{in}$ with respect to the surface normal [Fig. \ref{fig:configuration}(a)], 

\begin{equation}
\label{Eq:E_inc}
E_{x,1}^{\mathrm{inc}}(y,z) = E_\mathrm{in}e^{jk_1\textcolor{black}{z}\cos\theta_\mathrm{in}}e^{jk_{t_{0,1}}y}.
\end{equation}

As outlined in Section \ref{sec:introduction}, our objective is to design a MG that achieves anomalous reflection of an incident wavefront from an angle $\theta_\mathrm{in}$ to an angle $\theta_\mathrm{out}$ with maximum diffraction efficiency, while effectively mitigating absorption due to substrate and conductor loss. As shall be discussed in Section \textcolor{black}{\ref{subsec:prototype_design}}, using a single meta-atom per period as in \cite{Rabinovich2018} would generally not be sufficient when significant substrate loss exists, and leads to reduced performance \cite{diker2023low}. 
Thus, additional meta-atoms per period should be introduced, thereby increasing the available degrees of freedom and providing additional design flexibility. Keeping in mind our goal of a low-cost solution, we consider here the next-simplest MG configuration, featuring two meta-atoms per period (with impedance per unit length $\tilde{Z}_{1}$ and $\tilde{Z}_{2}$) separated by a distance $d$ and co-located on the plane $z=-h$, thereby enabling a straightforward single-layer PCB implementation [Fig. \ref{fig:configuration}(b)].

Considering this configuration, we commence by evaluating the fields scattered off the construct as a function of the MG parameters, 
building upon the analytical framework developed in \cite{Rabinovich2018,Popov2018,8892735}, 
with possibly lossy substrate $\varepsilon_2\in\mathbb{C}$. From the perspective of an observer in medium 1, $z<-h$, the total field is composed of two contributions. The first corresponds to the external fields, expressed as a superposition of the field emanating from the plane-wave excitation source (forward wave) and the field reflected from the grounded dielectric slab (backward wave) in the absence of the MG conductors, 
\begin{equation}
\label{Eq:E_ext}
        E_{x,1}^\mathrm{ext}(y,z) = E_\mathrm{in}(e^{-j\beta_{0,1}z} + R_{0}e^{j\beta_{0,1}(z+2h)})e^{jk_{t_{0,1}}y}
\end{equation}
where the transverse and longitudinal wavenumbers of the fundamental mode in medium 1 are given, respectively, as $k_{t_{0,1}} = k_{1}\sin\theta_\mathrm{0,1}$, $\beta_{0,1} = k_{1}\cos\theta_\mathrm{0,1}$, and the associated reflection coefficient is $R_{0} = (j\gamma_{0}\tan\beta_{0,2}h - 1)/(j\gamma_{0}\tan\beta_{0,2}h + 1)$\, with $\gamma_{0} = \zeta_{2}\beta_{0,1}/(\zeta_{1}\beta_{0,2})$.

The second contribution is related to the fields generated by the currents $I_{1}$ and $I_{2}$ induced within the loaded strips in response to the external field. These secondary fields can be effectively formulated as a series of laterally offset zeroth-order Hankel functions of the second kind (2D Green's functions), associated with the fields produced by the periodic array of thin electric line sources formed at $(y_{(1,n)},z)=(n\Lambda,-h)$ and $(y_{(2,n)},z)=(d+n\Lambda,-h)$ (carrying currents $I_1$ and $I_2$, respectively).
By employing the Poisson summation formula and exploiting the periodic nature of the structure, the Hankel function can be further represented as a superposition of discrete FB modes \cite{tretyakov2003analytical}, reading \cite{Rabinovich2018,8892735}
\begin{equation}
\label{Eq:E_grid}
\begin{split}
        &E_{x,1}^\mathrm{grid}(y,z) = \\
        &\,\, -\!\frac{k_{1}\zeta_{1}}{2\Lambda}\!\!\!\!\!\sum_{m=-\infty}^{\infty}\!\!\!\frac{T_{m}e^{-j\beta_{m,1}(z+h)}}{\beta_{m,1}}e^{jk_{t_{m,1}}y}\!\left(\!I_{1}\!+\! I_{2}e^{-jk_{t_{m,1}}d}\!\right)
\end{split}
\end{equation}
where the transverse ($k_{t_{m,i}}$) and longitudinal ($\beta_{{m,i}}$) wavenumbers of the $m$th FB mode in the $i$th medium are given by
\begin{equation}
\label{Eq:k&beta}
\begin{split}
        & k_{t_{m,i}} = k_{1}\sin\theta_\mathrm{in}+\frac{2\pi m}{\Lambda} = k_{i}\sin\theta_\mathrm{m,i}\\
        &\beta_{m,i} = \sqrt{k_{i}^2-k_{t_{m,i}}^2} = k_{i}\cos\theta_\mathrm{m,i}
\end{split}
\end{equation}
and the reflection ($R_{m}$) and transmission ($T_{m}$) coefficients 
of the $m$th FB mode, determined 
by imposing the continuity conditions of the tangential electric and magnetic field components on $z=-h$ and the vanishing of the tangential electric field on $z=0$, defined as\footnote{Note that the parameters in \eqref{Eq:E_ext} applicable to the external fields are simply the private case of \eqref{Eq:k&beta} and \eqref{Eq:R&T} as relevant to the fundamental FB harmonics ($m=0$).}
\begin{equation}
\label{Eq:R&T}
\begin{split}
        & R_{m} = \frac{j\gamma_{m}\tan\beta_{m,2}h - 1}{j\gamma_{m}\tan\beta_{m,2}h + 1}\\ 
        & T_{m} = 1 + R_{m} = \frac{2j\gamma_{m}\tan\beta_{m,2}h}{j\gamma_{m}\tan\beta_{m,2}h + 1}
\end{split}
\end{equation}
with the wave impedance ratio defined as
\begin{equation}
\label{Eq:w_imp_r}
\begin{split}
        \gamma_{m} = \frac{\zeta_{2}\beta_{m,1}}{\zeta_{1}\beta_{m,2}}.
\end{split}
\end{equation}
The total fields are the sum of the external and the grid generated fields 
\begin{equation}
\label{Eq:Tot_fields}
        E_{x,1}^\mathrm{tot}(y,z) = E_{x,1}^\mathrm{grid}(y,z) + E_{x,1}^\mathrm{ext}(y,z)
\end{equation}
Similarly, the fields inside the dielectric ($i=2$) are given by
\begin{equation}
\label{Eq: SubsFields}
\begin{split}
    &E_{x,2}^\mathrm{ext}(y,z) = -E_\mathrm{in}T_{0}\frac{\sin{(\beta_{0,2}z)}}{\sin{(\beta_{0,2}h)}}e^{j\beta_{0,1}h}e^{jk_{t_{0,2}}y}\\
    &E_{x,2}^\mathrm{grid}(y,z) = \\
    &\,\, -\!\frac{k_{1}\zeta_{1}}{2\Lambda}\!\!\!\!\!\sum_{m=-\infty}^{\infty}\!\!\!\frac{T_{m}}{\gamma_{m}\beta_{m,2}}\frac{\sin{(\beta_{m,2}z})}{\sin{(\beta_{m,2}h)}}e^{jk_{t_{m,2}}y}\!\left(\!I_{1}\!+\! I_{2}e^{-jk_{t_{m,2}}d}\!\right)\\
    &E_{x,2}^\mathrm{tot}(y,z) = E_{x,2}^\mathrm{grid}(y,z) + E_{x,2}^\mathrm{ext}(y,z)
\end{split}   
\end{equation}

\subsection{Induced currents}
\label{subsec:induced_currents}
We recall that our goal herein is to devise a MG for anomalous reflection, deflecting effectively the incoming power from $\theta_\mathrm{in}$ towards the desired angle $\theta_\mathrm{out}$ despite the reliance on \textcolor{black}{a} low-cost (lossy) substrate. Thus, as a first step, we fix the grid periodicity as to allow two propagating FB harmonics, 
$m = 0$ and $m = -1$ (specular and anomalous reflection, respectively), corresponding to the prescribed angles, namely
\begin{equation}
\label{Eq:gridPer}
\begin{split}
        \Lambda= \lambda/|\sin\theta_\mathrm{in} - \sin\theta_\mathrm{out}|
\end{split}
\end{equation}
In the spirit of low-cost designs, to yield the simplest MG (requiring the smallest number of degrees of freedom), we further stipulate that $\theta_\mathrm{in}$ and $\theta_\mathrm{out}$ be chosen such that they correspond to the only possible radiation channels, with all other higher-order modes ($m<-1$ or $m>0$) being evanescent \cite{ra2017metagratings, Rabinovich2018}.

As noted in Section \ref{subsec:scattered_fields} and shall be discussed also later in Section \ref{subsec:prototype_design}, 
we attempt to overcome the increased substrate loss by utilizing two meta-atoms per period. In other words, we aim to meet the design goals by judiciously tuning the four available MG degrees of freedom: the substrate thickness $h$, the interlement spacing $d$, and the load impedances-per-unit-length $\tilde{Z}_{1}$ and $\tilde{Z}_{2}$. 

The impact of these degrees of freedom on the scattered fields \eqref{Eq:Tot_fields}
is manifested by the currents induced within the grid, $I_{1}$ and $I_{2}$. For given $h$, $d$, $\tilde{Z}_{1}$, and $\tilde{Z}_{2}$, these can be evaluated via Ohm's law \cite{tretyakov2003analytical, Rabinovich2018}, 
\begin{equation}
\begin{split}
    &E_{x,1}^\mathrm{tot}(y\rightarrow0,z\rightarrow-h) = \tilde{Z}_{1}I_{1}\\ &E_{x,1}^\mathrm{tot}(y\rightarrow d,z\rightarrow-h) = \tilde{Z}_{2}I_{2} 
\end{split}
\end{equation}
 
leading to the following set of 
equations, 
\begin{equation}
\label{Eq:I0_I1}
\left\{
\begin{aligned}
\begin{split}
   \tilde{Z}_{1}I_{1} = & \,\, E_{in}(1 + R_{0})e^{j\beta_{0,1}h} \\
   &  \!-\! \frac{k_{1}\zeta_{1}}{2\Lambda}\!\!\sum_{m=-\infty}^{\infty}\!\!\!\frac{I_{1}R_{m} + I_{2}T_{m}e^{-jk_{t_{m,1}}d}}{\beta_{m,1}} \\
   &  -\!\frac{k_{1}\zeta_{1}}{2}I_{1}\!\left\{\!\!\!\begin{array}{l}
    \frac{1}{\Lambda\beta_{0,1}}\!+\!\frac{j}{\pi}\log{\frac{\Lambda}{2\pi r_\mathrm{eff}}}
   \\ \!+\!\frac{j}{2\pi}\sum_{\substack{m=-\infty \\ m \neq 0}  }^{\infty}\left(\frac{2\pi}{\Lambda\alpha_{m,1}}-\frac{1}{|m|}\right)\end{array}\!\!\!\!\right\} \\ \\
    \tilde{Z}_{2}I_{2} = & \,\, E_{in}(1 + R_{0})e^{j\beta_{0,1}h}e^{jk_{t_{0,1}}d} \\
   &  \!-\! \frac{k_{1}\zeta_{1}}{2\Lambda}\!\!\sum_{m=-\infty}^{\infty}\!\!\!\frac{I_{1}T_{m}e^{jk_{t_{m,1}}d} + I_{2}R_{m}}{\beta_{m,1}} \\
   &  -\!\frac{k_{1}\zeta_{1}}{2}I_{2}\!\left\{\!\!\!\begin{array}{l}
    \frac{1}{\Lambda\beta_{0,1}}\!+\!\frac{j}{\pi}\log{\frac{\Lambda}{2\pi r_\mathrm{eff}}}
    \\ \!+\!\frac{j}{2\pi}\sum_{\substack{m=-\infty \\ m \neq 0}  }^{\infty}\left(\frac{2\pi}{\Lambda\alpha_{m,1}}-\frac{1}{|m|}\right)\end{array}\!\!\!\!\right\}
\end{split}
\end{aligned} \right.
\end{equation}
where $\beta_{m,1} = -j\alpha_{m,1},\, (\alpha_{m,1} \geq 0)$, and $r_\mathrm{eff} = \frac{w}{4}$ is the effective strip radius based on the flat wire approximation \cite{Barkeshli2004}. 
Solving \eqref{Eq:I0_I1} for $I_1$ and $I_2$ with the relevant 
$\tilde{Z}_{q} = \tilde{R}_{q} + j\tilde{X}_{q}$ and $\varepsilon_\mathrm{2} = \varepsilon'_\mathrm{2,r}\varepsilon_0(1-j\tan\delta)$ (conductor and substrate loss manifested by the the resistive part of the load, $\tilde{R}_{q}\in\mathbb{R}$, and the dielectric loss tangent, $\tan\delta\in\mathbb{R}$), 
 one can ascertain the spatial distribution of the fields everywhere in space via substitution in (\ref{Eq:E_inc})-(\ref{Eq: SubsFields}),
  realistically accounting for the potential dissipation mechanisms within the MG. 

\subsection{Anomalous reflection MG}
\label{subsec:anomalous_reflection}
Given the scattering problem formulation in Sections \ref{subsec:scattered_fields} and \ref{subsec:induced_currents}, the process of obtaining effective anomalous reflection translates to maximizing the coupling efficiency between the incident wave and the $m=-1$ FB harmonics. Considering (\ref{Eq:E_inc})-(\ref{Eq:Tot_fields}), 
this anomalous reflection efficiency, $\eta_{-1}$, is given by the ratio of the power in the reflected $m=-1$ FB mode to the incident power,
\begin{equation}
\label{Eq:anomalEff}
        \eta_{-1} = \frac{1}{4} \left( \frac{\zeta_{1}}{\Lambda} \right)^2 \frac{|T_{-1}|^2}{\cos\theta_\mathrm{in}\cos\theta_\mathrm{out}} \left| \frac{I_{1} + I_{2}e^{-jk_{t_{-1,1}}d}}{E_\mathrm{in}} \right|^2,
\end{equation}
while the fraction of power $\eta_{0}$ undergoing specular reflection (coupling to the fundamental $m=0$ FB harmonic) can be quantified via
\begin{equation}
\label{Eq:specEff}
        \eta_{0} = |R_{0}|^2\Bigg|1-\frac{\zeta_{1}}{2\Lambda}\frac{1 + R_{0}}{R_{0}}\frac{e^{-j\beta_{0,1}h}}{\cos\theta_\mathrm{in}}\frac{I_{1} + I_{2}e^{-jk_{t_{0,1}}d}}{E_\mathrm{in}}\Bigg|^2.
\end{equation}
Since according to the FB theorem in our configuration, radiated power can only be coupled to the $m=-1$ and $m=0$ modes (see discussion around \eqref{Eq:gridPer}), the relative absorption (loss) in the resultant MG is given by
\begin{equation}
\label{Eq:loss}
    \eta_\mathrm{loss} = 1 - \eta_\mathrm{-1} - \eta_\mathrm{0},
\end{equation}
following global power conservation.

Equations \eqref{Eq:E_inc}-\eqref{Eq:loss} establish a complete analytical model relating a given MG configuration (Fig. \ref{fig:configuration}) to the fields scattered off it. Therefore, in order to realize a MG for effective anomalous reflection, we may harness this formalism to find the optimal set of degrees of freedom $(h,d,\tilde{Z}_{1},\tilde{Z}_{2})$ that maximizes $\eta_{-1}$ of \eqref{Eq:anomalEff}. The last stage of the synthesis procedure involves replacing the obtained distributed impedance values $\tilde{Z}_{1},\tilde{Z}_{2}$ with actual printed copper traces [featuring capacitor widths $W_{1}$ and $W_{2}$, \textit{cf.} Fig. \ref{fig:configuration}(c)]. This step, which will be discussed in great detail in Section \ref{subsec:dielectric_loss}, 
finalizes the design process, yielding a detailed practical PCB layout, ready for fabrication.

\section{Results and Discussion}
\label{sec:results}
The key aspects distinguishing this work from prior reports on beam deflecting MGs relate to the utilization of a low-cost lossy substrate. Thus, the current section is written with two main goals in mind: demonstrating how proper incorporation of conductor and substrate loss into the analytical model may still enable synthesis of effective anomalous reflection MG, verifying our theoretical derivation (Sections \ref{subsec:dielectric_loss} and \ref{subsec:prototype_design}, including an experimental validation in Section \ref{subsec:experiment}); and shedding light on the dominant loss mechanisms in this configuration and their contribution to the overall performance, providing essential insight for current and future applications (Section \ref{subsec:circuit_model}).

\subsection{Introducing dielectric loss}
\label{subsec:dielectric_loss}
Correspondingly, to demonstrate and validate the semianalytical design methodology, we focus on devising a PCB-based anomalous reflection MG prototype, utilizing FR4 as the dielectric substrate \textcolor{black}{(}denoted in Fig. \ref{fig:configuration} as medium 2\textcolor{black}{)}. As expected, such a low-cost choice usually exhibits non-negligible losses, characterized by a typical value of $\varepsilon_{2} = 4.4\varepsilon_0(1 - 0.02j)$ \cite{667012}. 
Our goal is to design the MG as to achieve efficient anomalous reflection, redirecting an incoming wave from $\theta_\mathrm{in}=10^\circ$ towards a given $\theta_\mathrm{out}$ at $f = 20$ GHz. 

In contrast to common MGs for diffraction engineering \cite{epstein2017unveiling,Rabinovich2018,Popov2018, xu2020dual,8892735,Xu2021}, in which the reliance on low-loss dielectrics and highly conductive metals allows one to ignore power dissipation as a zeroth-order approximation within the main synthesis procedure (and consider it later as a small perturbation), herein the inherent lossy nature of the substrate must be taken into account from the beginning of the design to yield meaningful results \cite{diker2023low}. 
Correspondingly, the design process should start with a characterization of the effective load impedance achievable with the chosen printed capacitor geometry (Fig. \ref{fig:configuration}), fully accounting for all possible dissipation factors. This would yield a look-up table identifying the range of load impedances $\tilde{Z}_{1}(W_1), \tilde{Z}_{2}(W_2)$ realizable based on the chosen low-cost substrate (FR4 in our prototype) and the proposed meta-atom configuration [Fig. \ref{fig:configuration}(c)], allowing later to reliably devise practical MGs in which desired diffraction efficiency is maximized over absorption.

To thus establish the relationship between the effective load impedance $\tilde{Z}_q$ and the capacitor width $W_q$, we 
employ the method outlined in \cite{Popov2019}. Specifically, we define in Ansys HFSS an auxiliary MG configuration consisting of a single loaded strip meta-atom per period, featuring periodicity\footnote{\label{foot:Lambda'}Considering the effective lumped (quasistatic) load impedance as an intrinsic property of the deeply subwavelength load geometry, we may choose $\Lambda'$ arbitrarily, in principle, as long as mutual coupling between laterally adjacent meta-atoms (along $y$) remains dominated by the dipole moment interaction (matching the uniform distributed impedance approximation)\textcolor{black}{.}} of $\Lambda'$ [Fig. \ref{fig:ExtrConfig1andLUT}(a)]. 
The load (printed capacitor) has a width of $W$, and is situated at the air/dielectric interface formed by a $h'$-thick PEC-backed dielectric substrate\footnote{\label{foot:h'}Similar to Footnote \ref{foot:Lambda'}, to retrieve the effective load impedance properly, we may, in principle, choose $h'$ arbitrarily. Nonetheless, we have noticed that certain combinations of $\Lambda'$ and $h'$ that correspond to resonant configurations (featuring very strong coupling to evanescent modes), may lead to diverging modal reflection coefficients $R_m$ in (\ref{Eq:R&T})
and distort the retrieval procedure, and should be avoided.
} ($\varepsilon=\varepsilon_\mathrm{sub}\in\mathbb{C}$), reproducing the near-field environment experienced by the meta-atom in the ultimate MG [Fig. \ref{fig:configuration}(a)]. 

The structure is illuminated by a normally incident plane wave, 
and its fundamental FB mode reflection coefficient $S_{11}$, corresponding in our formalism to
\begin{equation}
    S_{11} = [E_{x,1}^\mathrm{ext,R}(y,z) + E_{x,1}^\mathrm{grid,m=0}(y,z)]/E_{x,1}^\mathrm{in}(y,z)
    \label{Eq:S11}
\end{equation}
is \emph{recorded} by the full-wave solver at $z=-2\lambda$, for a given $W$. In \eqref{Eq:S11}, $E_{x,1}^\mathrm{ext,R}(y,z)$ stands for the external field reflected \textcolor{black}{off the} configuration in the absence of the loaded wires [the second term in (\ref{Eq:E_ext}) for the configuration of Fig. \ref{fig:configuration}(a)]\textcolor{black}{;} $E_{x,1}^\mathrm{grid,m=0}$ is the fundamental FB mode contributed by the primitive loaded-wire grid of this particular setup, corresponding to the zeroth order $m=0$ of (\ref{Eq:E_grid})\textcolor{black}{;} and the incident field for this scenario is $E_{x,1}^\mathrm{in}(y,z)=E_\mathrm{in}e^{-jkz}$. 
In the next step, the current $I$ effectively induced on the loaded wire in this configuration (i.e., for the current $W$) is calculated based on \eqref{Eq:E_inc}-\eqref{Eq:Tot_fields}, \eqref{Eq:S11}, and the recorded reflection coefficient as defined in \cite{Popov2019},
\begin{equation}
\label{Eq:Current_S11}
    I = \frac{2\Lambda \textcolor{black}{E_\mathrm{in}}(R_{0}e^{j2\beta_{0,1}(h-\lambda)})-S_{11}e^{j2\beta_{0,1}\lambda})}{\zeta_{1}(1+R_{0})e^{j\beta_{0,1}(h-2\lambda)}}
\end{equation}
in correspondence with the reduced formalism for a single meta-atom \cite{ra2017metagratings,Rabinovich2018}. 
Finally, the load impedance $\tilde{Z}_{T}$ is 
extracted based on Ohm's law $E_{x,1}^\mathrm{tot}(y\rightarrow0,z\rightarrow-h') = \tilde{Z}\textcolor{black}{_{T}}I$. After performing multiple iterations of the described calculation for various capacitor widths, we obtain the relation between the meta-atom geometry and the associated distributed load impedance $\tilde{Z}_{T}(W)=\tilde{R}_{T}(W)+j\tilde{X}_{T}(W)$. 

\begin{figure}[tbp]
    \centering
    \subfigure[]{\includegraphics[width=0.15\textwidth]{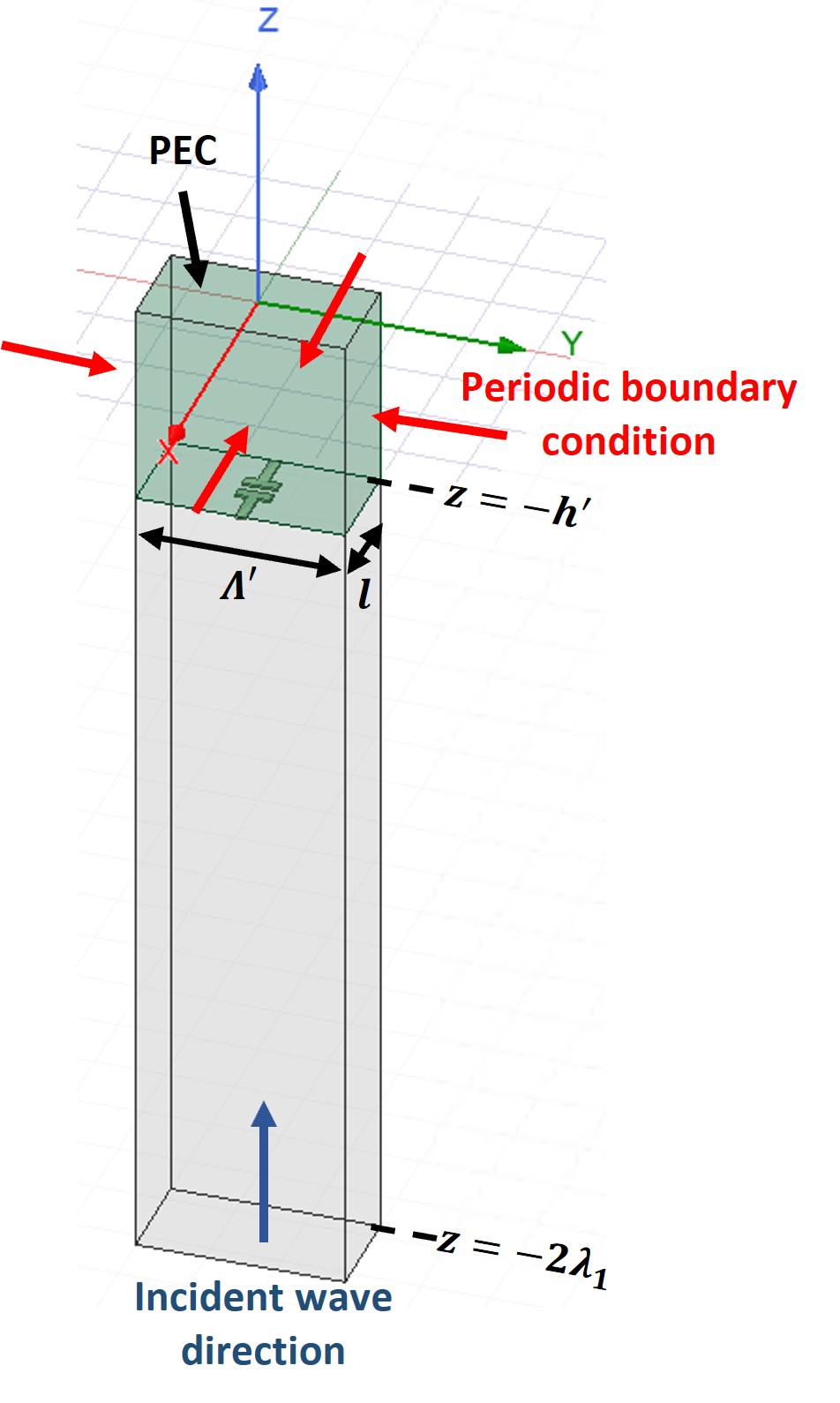}}
    \subfigure[]{\includegraphics[width=0.30\textwidth]{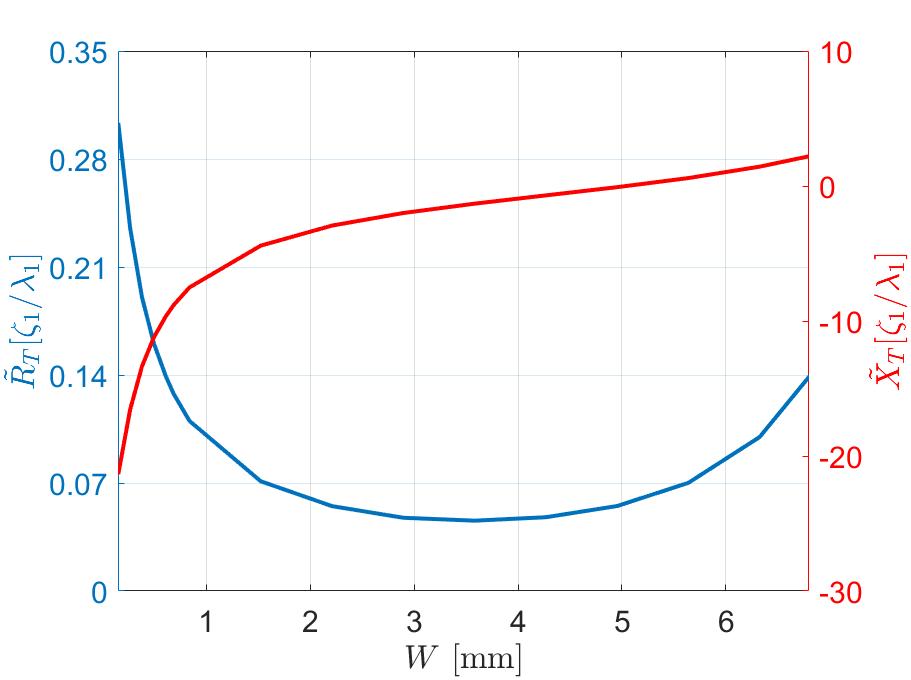}}
    \caption{(a) Physical configuration for the $\tilde{Z}\textcolor{black}{_T}(W)$ look-up table extraction (Section \ref{subsec:dielectric_loss}). A single meta-atom with printed capacitor load of width $W$ (swept in a full-wave solver) is placed in a ($l\times\Lambda'$)-periodic configuration, defined on a $h'$-thick PEC-backed FR4 substrate. The impedance extraction procedure involves recording the reflection coefficient $S_{11}$ at the plane $z=-2\lambda$ for a given $W$ using a full-wave solver, deducing the effective induced current $I$ via (\ref{Eq:Current_S11}), and finally evaluating the effective $\tilde{Z}_{T}(W)$ based on Ohm's law (similar to \eqref{Eq:I0_I1}). (b) The extracted relation between the meta-atom geometry and the associated distributed load impedance $\tilde{Z}_{T}(W)=\tilde{R}_{T}(W)+j\tilde{X}_{T}(W)$ \textcolor{black}{at the operating frequency $f=20$ GHz}.}
    \label{fig:ExtrConfig1andLUT}
\end{figure}

We executed the outlined procedure for the prototype configuration, featuring the printed copper capacitor of Fig. \ref{fig:configuration}(c) defined on FR4 substrate\footnote{We chose the substrate thickness and $y$ periodicity in the simulation setup of Fig. \ref{fig:ExtrConfig1andLUT} to be $h' = 0.315\lambda$ and $\Lambda' = 0.835\lambda$, respectively, in correspondence with Footnotes \ref{foot:Lambda'} and \ref{foot:h'}.}, using fabrication-compatible dimensions for the trace width, separation, and thickness, of $w=s=6$mil and $t=35\mu$m, respectively. The printed capacitors repeat periodically along $x$ every $l=\lambda/10$ [see Fig. \ref{fig:configuration}(b), (c)]\footnote{This subwavelength periodicity $l\ll\lambda$ justifies the modelling assumption of a uniform impedance per unit length loading and corresponding 2D analysis $\partial/\partial x\approx 0$ (Section \ref{sec:theory}).}, and we sweep the capacitor widths in the range $W\in[0.15, 6.8]$ mm. This yields the results presented in Fig. \ref{fig:ExtrConfig1andLUT}(b), specifying the effective (complex) load impedance $\tilde{Z}_{T}(W)=\tilde{R}_{T}(W)+j\tilde{X}_{T}(W)$ as a function of the meta-atom geometry. 

Beyond the practical implications of these results, needed to finalize the design procedure towards physical realization \textcolor{black}{(see end of Section \ref{subsec:anomalous_reflection})}, the obtained relation as depicted in Fig. \ref{fig:ExtrConfig1andLUT}(b) may provide insights into the loss mechanisms governing the meta-atom (and, correspondingly, the MG) performance. 
Notably, we observe that the resistive component of the effective impedance exhibits significantly higher values (at some points by an order of magnitude) 
compared to the ones associated with similar loads defined on low-loss substrates \cite{epstein2017unveiling,Popov2019}. Furthermore, these unexpectedly high values follow a rather uncommon trend as a function of $W$, characterized by a monotonic decrease to a point of minimum resistivity, followed by a subsequent monotonic increase. These seemingly peculiar features imply that a more thorough investigation and discussion are in place, which we conduct in the following\textcolor{black}{.}



\subsection{Equivalent circuit model}
\label{subsec:circuit_model}


\subsubsection{Physical considerations and general concept}
\label{subsub:physical_considerations}

\begin{figure}[tbp]
    \includegraphics[width=0.30\textwidth]{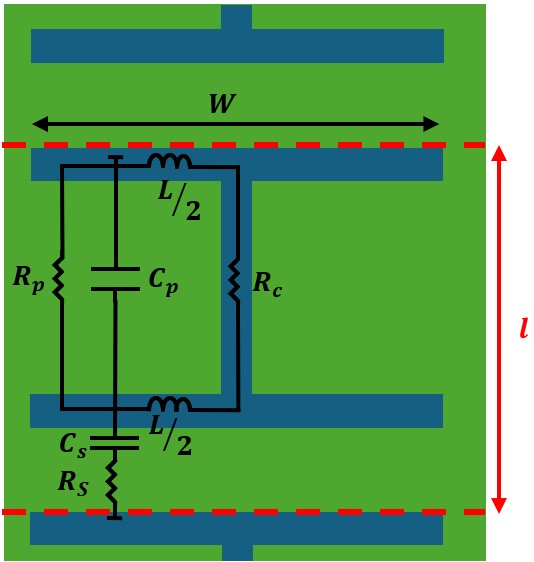}
    \centering
    \caption{Equivalent circuit of the MG printed load (corresponding to the effective distributed load impedance $\tilde{Z}_{T}$) of Fig. \ref{fig:Setup}(b) and (c). The printed load geometry incorporates two distinct RLC circuits: a series configuration and a parallel configuration (Section \ref{subsec:circuit_model}). The series RLC circuit (formed by $C_s$, $L$, $R_s$, and $R_c$) exhibits greater prominence within the lower range of $W$ values ($W<W_{s}$), while the parallel RLC circuit (formed by $C_p$, $L$, $R_p$, and $R_c$) becomes more dominant as W increases. 
    }
    \label{fig:EqvCirc}
\end{figure}

To provide insights and elucidate the mechanisms underlying the response observed in Section \ref{subsec:dielectric_loss}, their relation to the introduced lossy substrate, and their overall impact on the design, we develop in this section a suitable equivalent circuit model. 
We commence by analyzing the results presented in Fig. \ref{fig:ExtrConfig1andLUT}(b), 
describing the relation between the effective load (complex) impedance $\tilde{Z}_{T}$ and the printed capacitor width $W$ of Fig. \ref{fig:configuration}(c), as extracted from full-wave sweeps at the operating frequency $f=20$ GHz. 

Reviewing Fig. \ref{fig:ExtrConfig1andLUT}(b), one may observe that the reactive part of the impedance $\tilde{X}_{T}(W)=\Im\{\tilde{Z}_{T}(W)\}$ follows the characteristic behavior of a \emph{series LC} circuit, wherein distinct regions exhibit dominance by either the capacitive or inductive component. Specifically, the reactance is highly capacitive ($\tilde{X}_{T}\ll0$) for small capacitor widths $W\ll W_s=5.04$ mm 
(small capacitance); as this width increases, the capacitance increases until balance between inductive and capacitive response is achieved near the resonant width $W_{s}$; 
any further increase in $W$ beyond $W_s$ leads to residual inductive response, stemming from the elongation of the conductors \cite{owyang1958approximate,garg2024microstrip}. 
These main capacitive and inductive contributions per unit length appear as $\tilde{C}_{s}$ 
and $\tilde{L}$ 
in the equivalent circuit model we propose for the meta-atom, depicted in Fig. \ref{fig:EqvCirc}. 

We move next to examine the resistive part of the effective impedance $\tilde{R}_{T}(W)=\Re\{\tilde{Z}_{T}(W)\}$, as retrieved and presented in Fig. \ref{fig:ExtrConfig1andLUT}(b). As mentioned in Section \ref{subsec:dielectric_loss}, it is noticed that the effective resistance follows a non-monotonic trend with $W$, quite uncommon for loaded-wire PCB MGs relying on printed capacitors \cite{epstein2017unveiling, Popov2019, Wang2023}; in addition, the recorded resistances are significantly larger than previously reported ones for the configuration of Fig. \ref{fig:configuration} on \emph{low-loss} substrates. In the latter case, the effective resistive features of the load were naturally attributed solely to conductor loss, which is expected to be relatively insensitive to the capacitor width \cite{epstein2017unveiling}; it can be associated with the Ohmic losses along the main current flow path, symbolically denoted by $\tilde{R}_c$ in Fig. \ref{fig:EqvCirc}. However, in view of the additional dissipation channel introduced herein in the form of the lossy substrate, it would be reasonable to seek for additional sources for the rather unconventional $\tilde{R}_{T}(W)$ profile in the phenomena related to the non-negligible dielectric loss. 

\begin{figure}[tbp]
    \centering
    \subfigure[]{\includegraphics[width=0.34\textwidth]{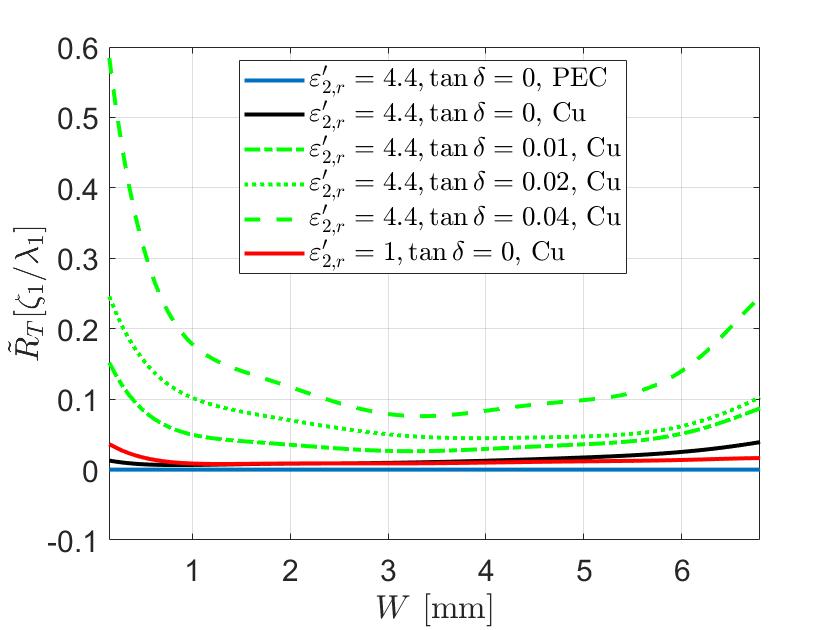}}
    \subfigure[]{\includegraphics[width=0.34\textwidth]{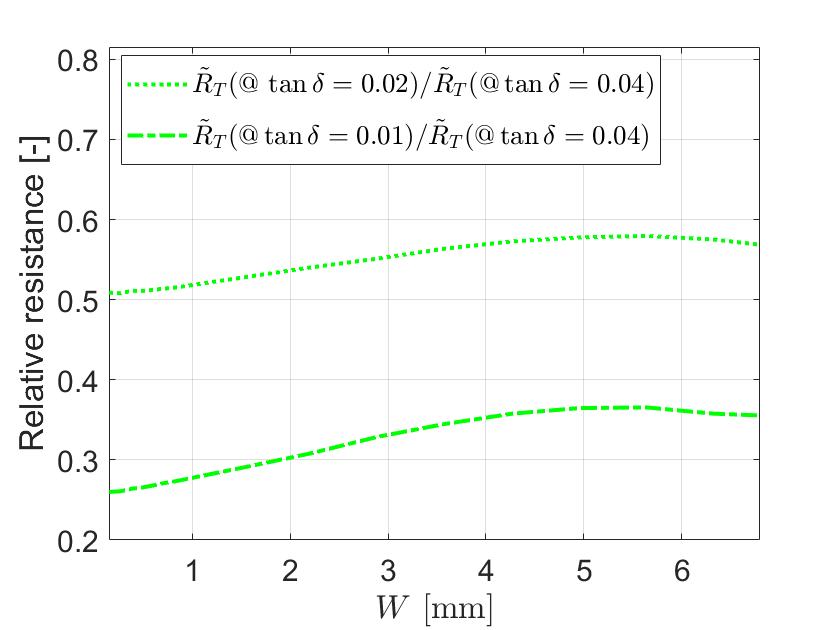}}
    \subfigure[]{\includegraphics[width=0.34\textwidth]{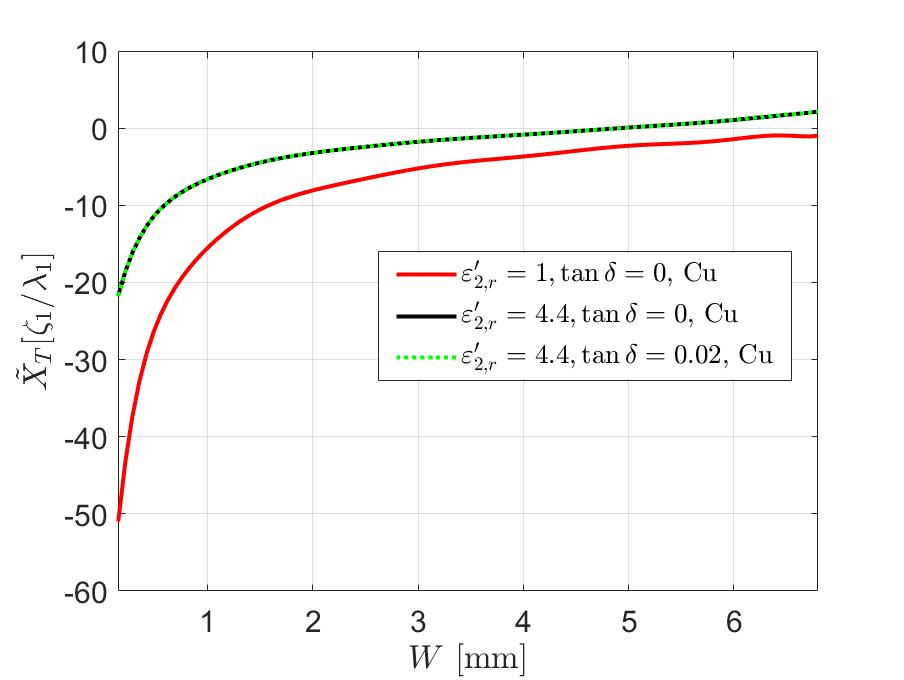}}
    \caption{Analysis of various loss factors contributing to the effective printed load \textcolor{black}{impedance $\tilde{Z}_{T}(W)=\tilde{R}_{T}(W)+j\tilde{X}_{T}(W)$} corresponding to Figs. \ref{fig:ExtrConfig1andLUT} and \ref{fig:EqvCirc}. (a) $\tilde{R}_{T}(W)$ as extracted from full-wave simulations following Section \ref{subsec:dielectric_loss} for the configuration in Fig. \ref{fig:ExtrConfig1andLUT} with different conductor properties, dielectric substrates, and loss level: lossless FR4 $\varepsilon_{2,\mathrm{r}}'=4.4$ as substrate with PEC (solid blue) or realistic finite-conductivity copper (solid black) as meta-atom conductors; FR4 substrates with loss tangent varying from $\tan\delta=0.01$ (dot-dashed green), \textcolor{black}{through} $\tan\delta=0.02$ (dotted green), to $\tan\delta=0.04$ (dashed green), considering realistic copper conductors; and a reference configuration with only conductor loss (solid red), featuring copper meta-atoms suspended in vacuum ($\varepsilon_{2,\mathrm{r}}'=1$). (b) Relative resistance between the extracted $\tilde{R}_{T}(W)$ from (a) for $\tan\delta=0.01$ (dash-dotted green) and $\tan\delta=0.02$ (dotted green) with respect to the case $\tan\delta=0.04$. For all cases, realistic copper is used and $\varepsilon_{2,r}'=4.4$. 
    (c) Similarly extracted $\tilde{X}_{T}(W)$ for representative scenarios from (a) with realistic copper conductors suspended in vacuum (solid red), or defined on a lossless (solid black) or realistic (dashed green) FR4. 
    }
    \label{fig:RX_tloss}
\end{figure}

Indeed, considering more closely Fig. \ref{fig:ExtrConfig1andLUT}(b) - focusing on the resistive part behavior - we may identify again 
two distinct regions: for very small capacitor widths (small capacitance) $W\ll W_s$ the effective resistance is high; it decreases gradually with increased $W$ until reaching the stationary point around $W\approx W_s$ 
(the resonant width of the equivalent series LC circuit discussed \textcolor{black}{earlier in} this subsection); and then starts to increase further as $W$ grows beyond $W_s$. 
The first region ($W<W_s$) may be interpreted within the same \emph{series RLC} circuit picture described earlier, if one may identify the physical mechanism that would cause reduction of the equivalent resistance when the capacitance increases ($W$ grows).

Correspondingly, to get more information about the causes to \textcolor{black}{this} unconventional profile of $\tilde{R}_{T}(W)$ we conduct several preliminary tests. Specifically, we run the impedance extraction methodology on several variations of the examined configuration of Fig. \ref{fig:configuration}, to allow discerning between the effects of the substrate and the conductors on the resistive and reactive components of the effective equivalent response of Fig. \ref{fig:ExtrConfig1andLUT}(b). First, we consider in Fig. \ref{fig:RX_tloss}(a) the extracted equivalent resistance $R\textcolor{black}{_T}$ as a function of the capacitor width $W$ when modifying the properties of the dielectric and conductors (geometry stays identical to Fig. \textcolor{black}{\ref{fig:ExtrConfig1andLUT}(a)}). As anticipated, for the trivial limiting case in which all loss mechanisms are suppressed in both dielectric and conductors (loss tangent set to $\tan\delta=0$ and the metal is considered as a PEC), the extracted resistance \textcolor{black}{vanishes} $\tilde{R}_{T}(W) = 0$ (solid blue), providing a basic "sanity check" to our procedure. Subsequently, we replace the \textcolor{black}{PEC} traces with realistic copper (conductivity of $5.8\times 10^7 \mathrm{\,S/m}$) 
 to isolate the contribution of the conductors to power dissipation ($\tan\delta$ is still $0$). The results demonstrate that for two different permittivity values, $\varepsilon_\mathrm{sub} = 4.4\varepsilon_0$ (solid black) and $\varepsilon_\mathrm{sub} = \varepsilon_0$ (solid red), the effective losses remained relatively constant at approximately $0.015\,[\zeta_{1}/\lambda_{1}]$ 
 (aligning with the values estimated in \cite{epstein2017unveiling}). Importantly, this finding indicates that, indeed, the conductor loss contribution $R_c$ may be considered to a good extent as independent in both the capacitor width and the substrate permittivity. Lastly, when increasing now the substrate loss tangent gradually to the nominal FR4 value of $\tan\delta=0.02$ and beyond (green curves in Fig. \ref{fig:RX_tloss}(a)), we observe the dramatic increase in the overall equivalent resistance, accompanied by the introduced non-monotonic dependency in the capacitor width $W$. The effects become more pronounced as we increase the loss tangent from $\tan\delta=0.01$ (dash-dotted green), through $\tan\delta=0.02$ (dotted green), to $\tan\delta=0.04$ (dashed green). Interestingly, the increase in effective resistance at low $W$ values is proportional to the loss tangent, and slightly deviates from this proportionality factor as $W$ increases [Fig. \ref{fig:RX_tloss}(b)]; this again implies an interplay between two mechanisms that dominate the equivalent circuit at the various $W$ regimes, and will be addressed in Section \ref{par:parallel_RLC}. 
In parallel, when examining the extracted reactance for the same scenarios\footnote{\textcolor{black}{Due} to the close proximity of the results obtained for different loss tangent values, only part of the cases considered in Fig. \ref{fig:RX_tloss}(a) \textcolor{black}{are} included in Fig. \ref{fig:RX_tloss}(c).} 
in Fig. \ref{fig:RX_tloss}(c), 
we find that $\tilde{X}_{T}(W)$ is clearly unaffected by dielectric loss, depending solely on the dielectric coefficient (real permittivity).

\paragraph{Series RLC regime}
\label{par:series_RLC}
Equipped with these initial observations, we begin our equivalent circuit construction by recalling again that the trend of $\tilde{X}_{T}(W)$ [Fig. \ref{fig:ExtrConfig1andLUT}(b), \ref{fig:RX_tloss}(c)] resembles the one expected of a series LC circuit. When metallic meta-atom features fitting such a circuit behavior are printed on a lossy dielectric substrate, the effective total impedance \textcolor{black}{(per unit length)} can be described by \cite{costa2012closed},
 \begin{equation}
\label{Eq:Ztot_LC_Series}
    \tilde{Z}_{s}(W;\omega) = \frac{1-\omega^2\tilde{L}(W)\tilde{C}_{s,v}(W)\varepsilon_\mathrm{eff}}{j\omega\tilde{C}_{s,v}(W)\varepsilon_\mathrm{eff}}
\end{equation}
where $\tilde{C}_{s,v}(W)$ is the effective capacitance \textcolor{black}{in} vacuum and $\varepsilon_\mathrm{eff}$ is the effective relative permittivity (dielectric constant) considering the different media surrounding the dielectric/air interface the printed capacitor is located at, $\varepsilon_\mathrm{1}$ and $\varepsilon_\mathrm{2}$
 \begin{equation}
\label{Eq:Permittivity_eff}
    \varepsilon_\mathrm{eff} = \varepsilon'_\mathrm{eff}[1-j(\tan\delta)_\mathrm{eff}]
\end{equation}
with $\varepsilon'_\mathrm{eff}=(\varepsilon'_\mathrm{2,r} + 1)/2$ and $(\tan\delta)_\mathrm{eff} = \varepsilon''_\mathrm{2,r}/(\varepsilon'_\mathrm{2,r} + 1)$. Under the assumption of small losses $\varepsilon'_\mathrm{2,r}\gg \varepsilon''_\mathrm{2,r}$, which is valid in our case, we may further simplify (\ref{Eq:Ztot_LC_Series}) and yield an explicit expression for $\Re{\{\tilde{Z}_{s}\}}=\tilde{R}_{s}$ and $\Im{\{\tilde{Z}_{s}\}}=\tilde{X}_{s}$, reading \cite{costa2012closed}

\begin{equation}
\label{Eq:$R_s$_definition}
\left\{
\begin{aligned}
\begin{split}
    &\tilde{R}_s(W;\omega) = \frac{(\tan\delta)_\mathrm{eff}}{\omega \tilde{C}_{s}(W)}= \frac{\tan\delta}{\omega \tilde{C}_{s}(W)}\frac{\varepsilon'_\mathrm{2,r}}{\varepsilon'_\mathrm{2,r}+1}\\
    &\tilde{X}_s(W;\omega) = \omega L(W)-\frac{1}{\omega \tilde{C}_{s}(W)}
\end{split}
\end{aligned} \right.
\end{equation}
where $\tilde{C}_{s}(W) = \tilde{C}_{s,v}(W) \varepsilon'_\mathrm{eff}$. The \textcolor{black}{second expression in \eqref{Eq:$R_s$_definition}} illustrates the emergence of the series resonant width $W=W_s$ when $\tilde{X}_s(W;\omega)\approx 0$, \textcolor{black}{while the first one} explains the behavior of $\tilde{R}_{T}(W)$ observed in Fig. \ref{fig:ExtrConfig1andLUT}(b) for short capacitor widths. According to this series RLC circuit prediction, $\tilde{R}_{s}(W)$ should be inversely proportional to $\tilde{C}_{s}(W)$ in this region, as indeed confirmed by the trend seen in Fig. \ref{fig:ExtrConfig1andLUT}(b): for $W<W_s$, $\tilde{R}_{T}(W)$ decreases as $\tilde{X}_{T}(W)$ increases ($|\tilde{X}_{T}(W)|$ decreases in this capacitively dominated region). In addition, \eqref{Eq:$R_s$_definition} predicts that $\tilde{R}_s(W)$ would grow linearly with $\tan\delta$, which is consistent with \textcolor{black}{the} equivalent resistance $\tilde{R}_{T}(W)$ extracted for various loss tangent values presented in Fig. \ref{fig:RX_tloss}(a), as well the proportionality factor calculated in Fig. \ref{fig:RX_tloss}(b) for very low capacitor widths. The residual differences for $W\ll W_s$ can be attributed to the very small conductor-loss-associated $\tilde{R}_c$ (black curve in \textcolor{black}{Fig.} \ref{fig:RX_tloss}(a)), which is only mildly dependent in $W$ (see discussion earlier in \ref{subsub:physical_considerations}) leading to an approximate relation $\tilde{R}_{T}(W)\approx \tilde{R}_c + \tilde{R}_s(W)$ in this regime.

\paragraph{Parallel RLC regime}
\label{par:parallel_RLC}
Reviewing the results in Section \ref{par:series_RLC}, we may quickly conclude that the formulated \textcolor{black}{resistive part} $\tilde{R}_s(W)$ of the \textcolor{black}{equivalent} series RLC \textcolor{black}{circuit} alone cannot account for the increase seen in Fig. \ref{fig:ExtrConfig1andLUT}(b) in the effective resistance after the series resonant width $W>W_s$, as the model \eqref{Eq:$R_s$_definition} predicts its contribution to the overall resistance to monotonically \emph{decrease} with the continued increase of the capacitance. Furthermore, Fig. \ref{fig:RX_tloss}(a) implies that also for these larger capacitor widths, the increase in effective resistance is heavily affected by the presence of dielectric loss (\textcolor{black}{green} curves therein), identifying it as an important factor in this regard.

To gather more information as to shed light on the behaviour in this inductive regime $\tilde{X}_{T}(W)>0$, we decided to examine 
the effective impedance variation over a broader range of capacitor widths. Employing thus the procedure outlined in Section \ref{subsec:dielectric_loss} on this extended range yields the results presented in Fig. \ref{fig:Extended_LUT}(a). 
Analyzing the resultant relation $\tilde{Z}(W)$, it is revealed that yet another resonant width emerges at $W=W_p= 9.144$ mm, \textcolor{black}{denoted by the rightmost dashed line}. 
In contrast to the first resonance at $W=W_s$ \textcolor{black}{(left dashed)}, where the effective total reactance vanishes, the second resonance leads to vanishing \emph{susceptance} (infinite impedance), which is characteristic of an equivalent \emph{parallel RLC} circuit \cite{svoboda2013introduction}. 

\begin{figure}[tbp]
    \centering
    \subfigure[]{\includegraphics[width=0.34\textwidth]{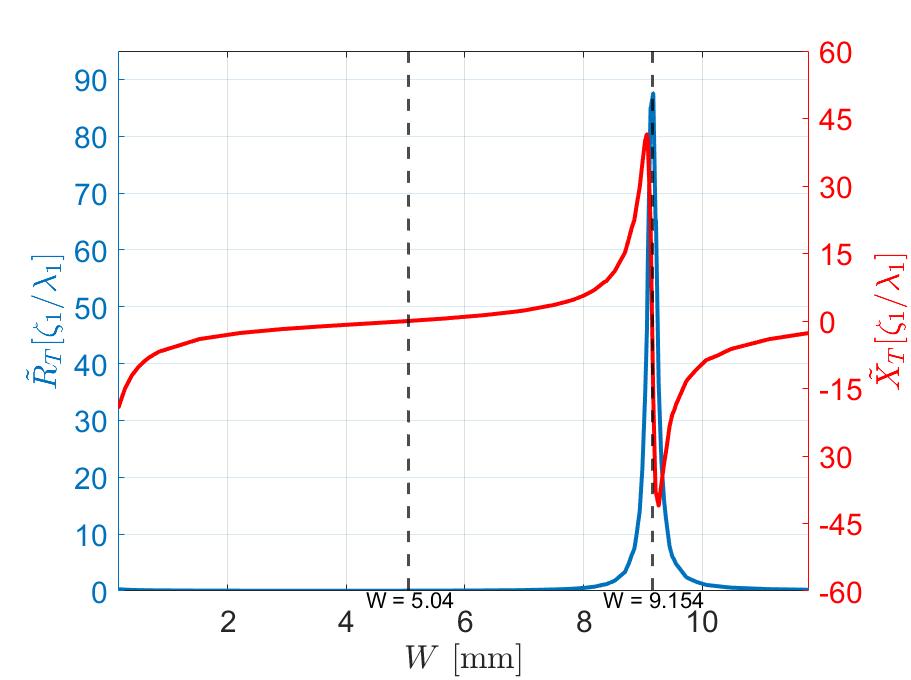}}
    \subfigure[]{\includegraphics[width=0.34\textwidth]{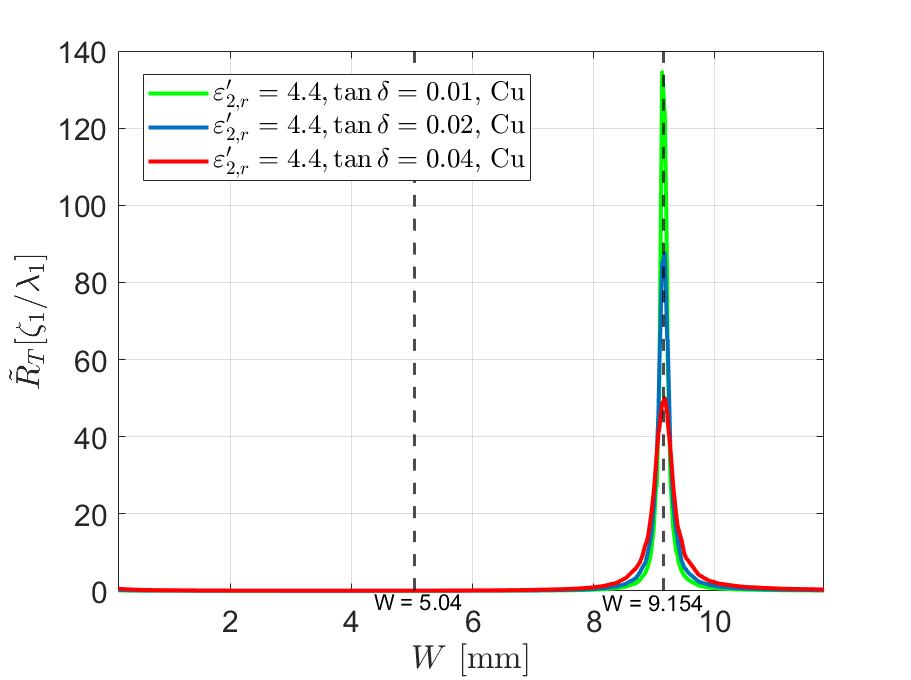}}
    \caption{\textcolor{black}{(a) Effective load resistance (blue solid curve, left $y$ axis) and reactance (red solid curve, right $y$ axis) as a function of capacitor widtth $W$, extracted from full-wave simulations following the procedure in Section \ref{subsec:dielectric_loss} for the configuration in Fig. \ref{fig:ExtrConfig1andLUT} in an extended range of capacitor widths (\textit{cf.} Section \ref{par:parallel_RLC}). (b) Effect of substrate loss tangent on the effective printed load resistance $\tilde{R}_{T}(W)$ extracted as in (a) for the same extended width. Results are shown for realistic copper meta-atom conductors and FR4 substrates $\varepsilon_{2,r}'=4.4$ with loss tangent varying from $\tan\delta=0.01$ (green), through realistic $\tan\delta=0.02$ (blue), to $\tan\delta=0.04$ (red).}}
    \label{fig:Extended_LUT}
\end{figure}

Reviewing the meta-atom configuration once more indicates the source for this parallel part of the equivalent circuit: parasitic \textcolor{black}{capacitance} $\tilde{C}_\mathrm{p}$ and resistance $\tilde{R}_\mathrm{p}$ (the latter due to leakage currents trough the substrate), induced by the finite distance between conductors from two adjacent capacitors along the $x$ \textcolor{black}{axis}. In Fig. \ref{fig:EqvCirc}, these contributions can be seen when considering the coupling between the two exterior legs of adjacent printed capacitors (forming a "dogbone" shape). 
As it turns out, the small parasitic capacitance $\tilde{C}_p$ along with the same line inductance $\tilde{L}$ due to the elongation of the capacitor legs discussed at the beginning of Section \ref{subsub:physical_considerations} reach this second balanced point as $W$ grows towards $W_p$, leading to increased effective resistance as expected from a parallel RLC \textcolor{black}{circuit} behaviour. The increase \textcolor{black}{of $R_T(W)$ observed} in Fig. \ref{fig:ExtrConfig1andLUT}(b) for $W>W_s$, thus, is actually the beginning of the transition of this dual-resonance circuit towards the second resonant width at $W=W_p$. 

\textcolor{black}{Corresdponginly, in analogy to the derivation in Section \ref{par:series_RLC} leading to \eqref{Eq:Ztot_LC_Series}, we may obtain an expression for the effective admittance \textcolor{black}{(inverse impedance per unit length)} of a metallic meta-atom with parallel RLC circuit characteristics when defined on a} lossy dielectric substrate\textcolor{black}{, reading}
\begin{equation}
\label{Eq:Ztot_LC_Parallel}
   \textcolor{black}{\tilde{Y}_{p}(W;\omega) =} \frac{1}{\tilde{Z}_{p}(W;\omega)} = \frac{1-\omega^2\tilde{L}(W)\tilde{C}_{p,v}(W)\varepsilon_\mathrm{eff}}{j\omega\tilde{L}(W)}
\end{equation}
\textcolor{black}{w}here $\tilde{C}_{p,v}(W)$ is the effective \textcolor{black}{parallel} capacitance \textcolor{black}{in} vacuum and $\varepsilon_\mathrm{eff}$ \textcolor{black}{is given by} (\ref{Eq:Permittivity_eff}). After rearranging (\ref{Eq:Ztot_LC_Parallel})} and separating \textcolor{black}{it to conductance $\Re{\{\tilde{Y}_{p}\}}=\tilde{G}_{p}=1/\tilde{R}_{p}$ and susceptance \textcolor{black}{$\Im{\{\tilde{Y}_{p}\}}=\tilde{B}_{p}$}}
, we obtain

\begin{equation}
\label{Eq:R_p_definition}
\left\{
\begin{aligned}
\begin{split}
    &\textcolor{black}{\tilde{G}_p(W;\omega)} = \omega \tilde{C}_{p}(W)(\tan\delta)_{\mathrm{eff}} = \frac{\varepsilon'_\mathrm{2,r}}{\varepsilon'_\mathrm{2,r}+1}\omega \tilde{C}_{p}(W)\tan\delta \\
    &\textcolor{black}{\tilde{B}_p(W;\omega)} = \omega \tilde{C}_{p}(W) - \frac{1}{\omega L(W)}
\end{split}
\end{aligned} \right.
\end{equation}
where $\tilde{C}_{p}(W) = \tilde{C}_{p,v}(W) \varepsilon'_\mathrm{eff}$ \textcolor{black}{(in analogy to the definitions after \eqref{Eq:Permittivity_eff})}. The expressions in (\ref{Eq:R_p_definition}) 
provide the "missing piece" to explain the overall effective impedance trends seen in Figs. \ref{fig:ExtrConfig1andLUT}(b), \ref{fig:RX_tloss}, and \ref{fig:Extended_LUT}\textcolor{black}{(a)}. When the capacitor width increases above the series resonant width $W>W_s$, the total impedance becomes dominated by the parallel RLC circuit branch (Fig. \ref{fig:EqvCirc}), as shall be discussed in greater detail in Section \ref{subsub:circuit_parameter_evaluation}. While in this intermediate inductive region $W_s<W<W_p$, the parallel capacitance $\tilde{C}_p(W)$ is still small (in the sense that its contribution to the susceptance is minor) and the parallel effective resistance (per unit length) can be approximated as
\begin{eqnarray}
\label{Eq:R_intermediate}
	 \textcolor{black}{\tilde{R}_{T}(W)\approx}\Re\{\tilde{Z}_p(W;\omega)\}\approx \omega^2 \tilde{L}(W)^2\textcolor{black}{\tilde{G}_p(W;\omega)} \nonumber \\
	 =\omega^3 \tilde{L}(W)^2\tilde{C}_{p}(W)\textcolor{black}{\frac{\varepsilon'_\mathrm{2,r}}{\varepsilon'_\mathrm{2,r}+1}}(\tan\delta)   
\end{eqnarray}
which explains both the increase seen in total effective resistance $\tilde{R}_{T}(W)$ as a function of $W$ in this region [Fig. \ref{fig:ExtrConfig1andLUT}(b)] due to the increasing inductance and capacitance, as well as the proportionality to $\tan\delta$ reflected in Fig. \ref{fig:RX_tloss}(a) and (b). On the other hand, as the capacitor width approaches the parallel resonant width $W_p$, the inductance and parallel capacitance balance one another, leading to peaking equivalent reactance $|X\textcolor{black}{_T}(W)|$ (minimal susceptance of the resonant parallel RLC circuit, \textit{cf.} Fig. \ref{fig:Extended_LUT}\textcolor{black}{(a)}). In this case, $\textcolor{black}{\tilde{R}_{T}(W)\approx}\Re\{\tilde{Z}_p(W;\omega)\}\approx\textcolor{black}{1/\tilde{G}_p(W;\omega)=}\tilde{R}_p(W;\omega)$ and we may use \eqref{Eq:R_p_definition} to assess the equivalent total resistance, peaking at $W=W_p$ as well. \textcolor{black}{In fact, the latter conclusion implies through \eqref{Eq:R_p_definition} that the total effective resistance at this parallel circuit resonance $W=W_p$ should be approximately \emph{inverse} proportional to the substrate loss tangent, as can indeed be deduced from Fig. \ref{fig:Extended_LUT}(b), where this dependency is recorded (extending the range of Fig. \ref{fig:RX_tloss}(a)).}

Overall, the identification of two resonant widths $W_s$ and $W_p$ corresponding to the equivalent circuit in Fig. \ref{fig:EqvCirc} enables proper physical interpretation of the effect of dielectric loss on the effective load impedance in our low-cost MG, seen in Fig. \ref{fig:Extended_LUT}. The discussion and corresponding formulation \eqref{Eq:Ztot_LC_Series}-\eqref{Eq:R_intermediate} clearly establish the crucial role the loss tangent and dielectric permittivity have in determining dissipation levels stemming from the combined series and parallel RLC circuits associated with our meta-atom, manifested in the equivalent total resistance trends.

\subsubsection{Circuit parameter evaluation}
\label{subsub:circuit_parameter_evaluation}
Once the general physical considerations have been discussed and qualitatively justified in Section \ref{subsub:physical_considerations}, we proceed to devise a systematic procedure for the assessment of the various equivalent circuit parameters established therein. Such an evaluation would allow for a quantitative validation of the proposed model, facilitating identification of the dominant mechanisms and their relative impact on the overall MG performance (and design).



Relying on the analysis in Section \ref{subsub:physical_considerations} leading to the equivalent circuit model in Fig. \ref{fig:EqvCirc}, we may formulate the total effective \textcolor{black}{distributed} load impedance $\tilde{Z}_{T}(W;\omega)$ of a meta-atom with \textcolor{black}{a} given $W$ at \textcolor{black}{the} frequency of operation $f=\omega/(2\pi)$ as 
\begin{equation}
\label{Eq:Gen_Imped}
\begin{split}
    \tilde{Z}_{T}(W;\omega)\!&=\!\tilde{R}_\mathrm{s}(W;\omega) + \tilde{X}_\mathrm{C_s}(W;\omega) + \\
   &\!\!\!\!\!\!\!\!\!\!\!\!+(\tilde{X}_\mathrm{L}(W;\omega) + \tilde{R}_\mathrm{c})||\tilde{X}_\mathrm{C_p}(W;\omega)||\tilde{R}_\mathrm{p}(W;\omega)
\end{split}
\end{equation}
where $\tilde{X}_\mathrm{C_s}(W;\omega)=1/\textcolor{black}{[}j\omega\tilde{C}_\mathrm{s}(W)\textcolor{black}{]}$, $\tilde{X}_\mathrm{C_p}(W;\omega) =1/\textcolor{black}{[}j\omega\tilde{C}_\mathrm{p}(W)\textcolor{black}{]}$ and $\tilde{X}_\mathrm{L}(W;\omega)=j\omega\tilde{L}(W)$. 
Correspondingly, we adopt very simplistic models for describing the detailed dependenc\textcolor{black}{ies} of these various circuit components in the capacitor width $W$ (the only degree of freedom in our meta-atom design), thus tying the abstract effective parameters to the actual trace geometry in the ultimate fabrication compatible PCB MG. Specifically, we assume that the capacitances (of both the \textcolor{black}{series} $\tilde{C}\textcolor{black}{_s}(W)$ and para\textcolor{black}{llel} $\tilde{C}_p(W)$ \textcolor{black}{circuit} capacitors) \textcolor{black}{grow linearly with} the capacitor width $W$ and the effective permittivity (as for conventional parallel plate capacitors \cite{epstein2017unveiling, Rabinovich2018} \cite[Eqs. \textcolor{black}{(7.78)-(7.80)}]{garg2024microstrip};
the effective inductance $L(W)$ \textcolor{black}{is proportional to the same} $W$\textcolor{black}{, acting in this context as the length of the conducting leg} \cite{owyang1958approximate};
the main conductor resistance $\tilde{R}_c$ is relatively independent of $W$ (current flows predominantly in the central conductor) and determined mostly by the metal conductivity\textcolor{black}{, the copper trace cross section, and the skin depth} \cite{epstein2017unveiling}; the resistance associated with the dielectric loss effect on the \textcolor{black}{series} capacitor impedance, $\tilde{R}_s\textcolor{black}{(W)}$, follows \eqref{Eq:$R_s$_definition}, i.e. linear with the loss tangent and inversely proportional to the \textcolor{black}{series} capacitance; and the resistance $\tilde{R}_p\textcolor{black}{(W)=1/\tilde{G}_p(W)}$ that manifests the impact of the same dielectric loss on the \textcolor{black}{parallel} capacitor impedance follows \eqref{Eq:R_p_definition}, inversely proportional to the parasitic capacitance \textcolor{black}{and loss tangent}. \textcolor{black}{Explicitly, we define}
\begin{align}
\label{Eq:CircuitParameters}
&\textcolor{black}{\tilde{C}_s(W)=\tilde{C}_{s,0}W+\tilde{C}_{s,1};\,\,\,\tilde{C}_p(W)=\tilde{C}_{p,0}W} \nonumber \\
&\textcolor{black}{\tilde{L}(W)=\tilde{L}_0W;\,\,\,\tilde{R}_c(W)\equiv\tilde{R}_c}
\end{align}
\textcolor{black}{where $\tilde{C}_{s,0}$, $\tilde{C}_{p,0}$, and $\tilde{L}_0$ are the proportionality factors for the series capacitance, parallel capacitance, and inductance, respectively; $\tilde{C}_{s,1}$ is the residual series capacitance due to the capacitive coupling between the two edges of "cut-wire" form of the meta-atom when $W\rightarrow0$; and $\tilde{R}_c$ is the average value of the main copper conductor resistance. Assessing the values of all these coefficients, in conjunction with \eqref{Eq:$R_s$_definition} and \eqref{Eq:R_p_definition}, would determine the entire behaviour of the equivalent circuit of Fig. \ref{fig:EqvCirc} as a function of $W$, and, in fact, also its complete frequency response (see discussion in Section \ref{subsub:full-wave_circuit_model}).}

Under these assumptions, the explicit expressions for the total resistive $\tilde{R}_T(W;\omega)=\Re\{\tilde{Z}_{T}(W;\omega)\}$ and reactive $\tilde{X}_T(W;\omega)=\Im\{\tilde{Z}_{T}(W;\omega)\}$ \textcolor{black}{components of the total effective distributed impedance of} \eqref{Eq:Gen_Imped} are given by 
\begin{equation}
\label{Eq:Imped}
\!\!\!\!\!\!\left\{
\begin{aligned}
\begin{split}
   & \tilde{R}_{T}(W;\omega)\!=\!\frac{\tilde{G}_{p\textcolor{black}{,T}}(W;\omega)}{\tilde{G}_{p\textcolor{black}{,T}}(W;\omega)^2+\tilde{B}_{p\textcolor{black}{,T}}(W;\omega)^2} + \frac{(\tan\delta)_\mathrm{eff}}{\omega \tilde{C}_\mathrm{s}(W)}\\
   & \tilde{X}_{T}(W;\omega)\!=\! \frac{\tilde{B}_{p\textcolor{black}{,T}}(W;\omega)}{\tilde{G}_{p\textcolor{black}{,T}}(W;\omega)^2+\tilde{B}_{p\textcolor{black}{,T}}(W;\omega)^2}
    - \frac{1}{\omega \tilde{C}_\mathrm{s}(W)}\\
\end{split}
\end{aligned} \right.
\end{equation}
where the equivalent \textcolor{black}{total} conductance $\tilde{G}_{p\textcolor{black}{,T}}$ and susceptance $\tilde{B}_{p\textcolor{black}{,T}}$ of the complete parallel circuit component as appears in Fig. \ref{fig:EqvCirc} are defined \textcolor{black}{by integrating the conductor loss $R_c$ into \eqref{Eq:R_p_definition}, reading} 
\begin{align}
\label{Eq:Gp_Bp_Imped}
&\tilde{G}_{p\textcolor{black}{,T}}(W;\omega)=\omega \tilde{C}_{p}(W)(\tan\delta)_{\mathrm{eff}}+\frac{\tilde{R}_\mathrm{c}}{\tilde{R}_\mathrm{c}^2+\left[\omega\tilde{L}(W)\right]^2} \nonumber \\
&\tilde{B}_{p\textcolor{black}{,T}}(W;\omega)=\frac{\omega\tilde{L}(W)}{\tilde{R}_\mathrm{c}^2+\left[\omega\tilde{L}(W)\right]^2} - \omega \tilde{C}_\mathrm{p}(W)
\end{align}

\textcolor{black}{We thus may commence evaluating the parameters of \eqref{Eq:CircuitParameters} from the full-wave results of Figs. \ref{fig:ExtrConfig1andLUT}, \ref{fig:RX_tloss}, and \ref{fig:Extended_LUT}.} The evaluation \textcolor{black}{procedure} is conducted at our designated operating frequency, $f=20$ GHz, considering an FR4 substrate 
$\varepsilon = 4.4\varepsilon_0(1 - 0.02j)$ as before 
(Section \ref{subsec:dielectric_loss}), relying on \eqref{Eq:Imped} and the insights from Section \ref{subsub:physical_considerations} \textcolor{black}{as follows:}

\begin{enumerate}[label=(\roman*)]
	\item \label{itm:Rc} \textcolor{black}{$\tilde{R}_c$:} The contribution of conductor \textcolor{black}{loss} 
	may be isolated by considering the effective resistance extracted from full-wave simulations when dielectric loss is absent ($\tan\delta=0$) and realistic copper is used to model the metallic traces. This corresponds to the solid black curve in Fig. \ref{fig:RX_tloss}(a), from which an average value of $\textcolor{black}{\tilde{R}}_c=0.015\,[\zeta_{1}/\lambda_{1}]$ may be estimated (\textcolor{black}{correlating well} with the ones estimated in \cite{epstein2017unveiling} for \textcolor{black}{generally similar frequencies and copper trace dimensions}), approximately independent of $W$.
	\item \label{itm:Cs} \textcolor{black}{$\tilde{C}_\mathrm{s,0},\,\tilde{C}_\mathrm{s,1}$: The series capacitance parameters are estimated from the region where the series RLC circuit is in its deeply capacitive regime $W\ll W_s$, prior reaching the series resonance condition and very far from the parallel resonance width $W_p$ (Figs. \ref{fig:ExtrConfig1andLUT}(a) and \ref{fig:ExtrConfig1andLUT}). In this range of $W$ values, the parallel capacitance is very small  $\omega \tilde{C}_{p}(W)\ll[\tilde{R}_c+\omega \tilde{L}(W)]/\{\tilde{R}_c^2+[\omega \tilde{L}(W)]^2\}$, and the current flowing through $\tilde{C}_p$ and $\tilde{R}_p(\propto1/\tilde{C}_p)$ may be neglected in \eqref{Eq:Gp_Bp_Imped}, leaving in the equivalent circuit of Fig. \ref{fig:EqvCirc} only the series RLC branch, reducing \eqref{Eq:Imped} to \eqref{Eq:Ztot_LC_Series} (up to an additional small $\tilde{R}_c$ in series). Moreover, since we are focusing on the deeply capacitive region of the series RLC circuit, $\omega\tilde{L}(W)\ll1/[\omega\tilde{C}_s(W)]$, we can neglect the inductance contribution as well, ultimately allowing approximation of the total equivalent resistance of \eqref{Eq:Imped} as $\tilde{R}_{T}(W;\omega)\approx \frac{(\tan\delta)_\mathrm{eff}}{\omega \tilde{C}_\mathrm{s}(W)}$ (where we considered $\tilde{R}_\mathrm{c} \ll \tilde{R}_\mathrm{s}$ based on the Fig. \ref{fig:RX_tloss}(a)). Subsequently, linear fitting $1/\tilde{R}_{T}(W)$ from Fig. \ref{fig:ExtrConfig1andLUT} (blue solid line) in this regime} $W\textcolor{black}{\in}[0.15,0.85]$ mm using the MATLAB \textcolor{black}{library routine \texttt{fminsearch}, yields the series capacitor coefficients of \eqref{Eq:CircuitParameters} as}  $\tilde{C}_\mathrm{s,0} = 1.096\textcolor{black}{\times10}^{-12}[\frac{\lambda_1\textcolor{black}{\cdot\mathrm{sec}}}{\zeta_1\textcolor{black}{\cdot\mathrm{mm}}}]\textcolor{black}{=43.608\,\mathrm{fF}}$ and $\tilde{C}_\mathrm{s,1} = 2.602\textcolor{black}{\times10}^{-13}[\frac{\lambda_1\textcolor{black}{\cdot\mathrm{sec}}}{\zeta_1}]\textcolor{black}{=10.353\,[\mathrm{fF}\cdot\mathrm{mm}]}$. 
	\item \label{itm:Cp} \textcolor{black}{$\tilde{C}_{p,0}:$ The parallel capacitance coefficient is found by enforcing the two (series and parallel) resonant conditions at $W=W_s$ and $W=W_p$, respectively. As the capacitor width grows from $W=0$ towards} the region around the \textcolor{black}{series} resonance \textcolor{black}{width} $W\rightarrow W_\mathrm{s}$\textcolor{black}{, the series and parallel capacitances increase, as well as the inductance (\textit{cf.} \eqref{Eq:CircuitParameters}). However, since this region is still quite far from the parallel resonant width, the total parallel susceptance of \eqref{Eq:Gp_Bp_Imped} is substantially larger than the conductance of the parallel RLC circuit, $\tilde{B}_{p,T}(W;\omega)\gg\tilde{G}_{p,T}(W;\omega)$. Considering that the inductance for such widths is already significant compared to the small copper resistance found in Step \ref{itm:Rc}, $\omega \tilde{L}(W) \gg \tilde{R}_\mathrm{c}$, we may approximate \eqref{Eq:Gp_Bp_Imped} as} $\tilde{B}_{p\textcolor{black}{,T}}(W;\omega)\approx\frac{1}{\omega \tilde{L}(W)} - \omega \tilde{C}_{p}(W)$ \textcolor{black}{around $W=W_s$}. Substituting \textcolor{black}{the last two results into \eqref{Eq:Imped} and demanding that the total effective reactance vanishes at the series resonant width, i.e.}  $\tilde{X}_{T}(W_\mathrm{s};\omega)=0$\textcolor{black}{, we may obtain a relation between the inductance and (series and parallel) capacitances at this width, reading} 
\begin{equation}
\label{Eq:L_W_s}
\tilde{L}(W_\mathrm{s})\approx\frac{1}{\omega^2(\tilde{C}_\mathrm{s}(W_\mathrm{s}) + \tilde{C}_\mathrm{p}(W_\mathrm{s}))} = \textcolor{black}{\tilde{L}}_{0}W_\mathrm{s},
\end{equation}	
\textcolor{black}{where the last equality stems from \eqref{Eq:CircuitParameters}. On the other hand, around the parallel resonant width $W=W_p$, the series capacitance grows further as to effectively short-circuit the series branch of the equivalent circuit (bottom of Fig. \ref{fig:EqvCirc}), $1/[\omega\tilde{C}_s(W)]\ll\frac{\tilde{B}_{p\textcolor{black}{,T}}(W;\omega)}{\tilde{G}_{p\textcolor{black}{,T}}(W;\omega)^2+\tilde{B}_{p\textcolor{black}{,T}}(W;\omega)^2}$. Correspondingly, enforcing the parallel resonant condition at this width $\tilde{B}_{p\textcolor{black}{,T}}(W_p;\omega)=0\Leftrightarrow\tilde{X}_{T}(W_\mathrm{p};\omega)=0$ requires through \eqref{Eq:CircuitParameters}-\eqref{Eq:Gp_Bp_Imped} that} 
\begin{equation}
\label{Eq:L_W_p}
\tilde{L}(W_\mathrm{p})\approx\frac{1}{\omega^2\tilde{C}_\mathrm{p}(W_\mathrm{p})} = \textcolor{black}{\tilde{L}}_{0}W_\mathrm{p},
\end{equation}	
\textcolor{black}{considering that $\omega \tilde{L}(W) \gg \tilde{R}_\mathrm{c}$ still holds due to the linear increase of the inductance from $W=W_s$ to $W=W_p$. Dividing \eqref{Eq:L_W_s} by \eqref{Eq:L_W_p} and substituting the relations \eqref{Eq:CircuitParameters}, we arrive at} $\tilde{C}_\mathrm{p,0}\textcolor{black}{=}\frac{W_\mathrm{s}\tilde{C}_\mathrm{s}(W\textcolor{black}{_s})}{W_\mathrm{p}^2 - W_\mathrm{s}^2}$\textcolor{black}{, which, using the values extracted in Step \ref{itm:Cs} for the series capacitance, yields} $\tilde{C}_\mathrm{p,0} = 4.991\textcolor{black}{\times10}^{-13}[\frac{\lambda_1\textcolor{black}{\cdot\mathrm{sec}}}{\zeta_1\textcolor{black}{\cdot\mathrm{mm}}}]\textcolor{black}{=19.859\,\mathrm{fF}}$. \textcolor{black}{Indeed, comparing the expected ratio $\tilde{C}_\mathrm{p,0}/\tilde{C}_\mathrm{s,0}\approx0.55$ based on the difference between the corresponding inter-strip distances ($s=6\,\mathrm{mil}=0.1524\,\mathrm{mm}$ and $l-s-2w=1.0428\,\mathrm{mm}$, respectively) and the complete elliptical integral ratio governing the equivalent capacitance of the coplanar strip waveguide \cite[Eq. (7.78)-(7.80)]{garg2024microstrip} resembling the scenario considered herein (Fig. \ref{fig:EqvCirc}), we observe good correspondence with the values $\tilde{C}_\mathrm{p,0}$ and $\tilde{C}_\mathrm{s,0}$ extracted in this step and in Step \ref{itm:Cs}, respectively.}
\item \label{itm:L0} \textcolor{black}{$\tilde{L}_0$:} Finally, \textcolor{black}{we evaluate the inductance coefficient via \eqref{Eq:L_W_p} and \eqref{Eq:CircuitParameters} with the parallel capacitance coefficient found in Step \ref{itm:Cp}, yielding} $\tilde{L}_{0} \textcolor{black}{=} 1.514\textcolor{black}{\times10}^{-12} [\frac{\zeta_1\textcolor{black}{\cdot\mathrm{sec}}}{\lambda_1\textcolor{black}{\cdot\mathrm{mm}}}]\textcolor{black}{=38.051\,\mathrm{[pH/(mm)^2]}}$.
\end{enumerate}

\begin{table}[t]
\centering
\caption{\textcolor{black}{Equivalent circuit parameters corresponding to Fig. \ref{fig:EqvCirc}, as extracted from full-wave simulations at $f=20$ GHz (Figs. \ref{fig:RX_tloss} and \ref{fig:Extended_LUT}) following the procedure laid out in Section \ref{subsub:circuit_parameter_evaluation}}.}
\textcolor{black}{\begin{tabular}{c c c c c} 
 \hline\hline
  $\tilde{R}_c[\Omega/\mathrm{mm}]$ & $\tilde{C}_{s,0}[\mathrm{fF}]$ & $\tilde{C}_{s,1}[\mathrm{fF}\cdot\mathrm{mm}]$ & $\tilde{C}_{p,0} [\mathrm{fF}]$ & $\tilde{L}_{0}[\mathrm{pH}/\mathrm{mm}^2]$\\ [0.7ex]
 \hline
0.377 & 43.608 & 10.353 & 19.859 & 38.051 \\
 \hline\hline
\end{tabular}}
\label{tab:CircuitParameters}
\end{table}
 
\textcolor{black}{Table \ref{tab:CircuitParameters} summarizes the estimated values of all circuit parameter coefficients, enabling through \eqref{Eq:CircuitParameters}-\eqref{Eq:Gp_Bp_Imped} evaluation of the complete effective load characteristics as a function of the load capacitor width. Notably, the evaluation procedure relies on simplified dependencies \eqref{Eq:CircuitParameters} and physically intuitive steps, focusing only at specific properties of the devised equivalent circuit of Fig. \ref{fig:EqvCirc} deduced from isolated conductor loss simulations (Step \ref{itm:Rc}), linear trend in the capacitive regime of the series RLC circuit (Step \ref{itm:Cs}), and the widths in which the two (series and parallel) RLC circuit resonant conditions are reached (Steps \ref{itm:Cp} and \ref{itm:L0}). Despite this simplicity, as shall be shown in Section \ref{subsub:full-wave_circuit_model} to follow, the circuit model with the same evaluated parameters shows an impressive ability to capture the meta-atom response beyond the specific operating conditions from which they were estimated, providing very good predictions for the load properties across a band of frequencies as well.}

\subsubsection{Full-wave validation}
\label{subsub:full-wave_circuit_model}
\begin{figure}[tbp]
    \centering
    \subfigure[]{\includegraphics[width=0.38\textwidth]{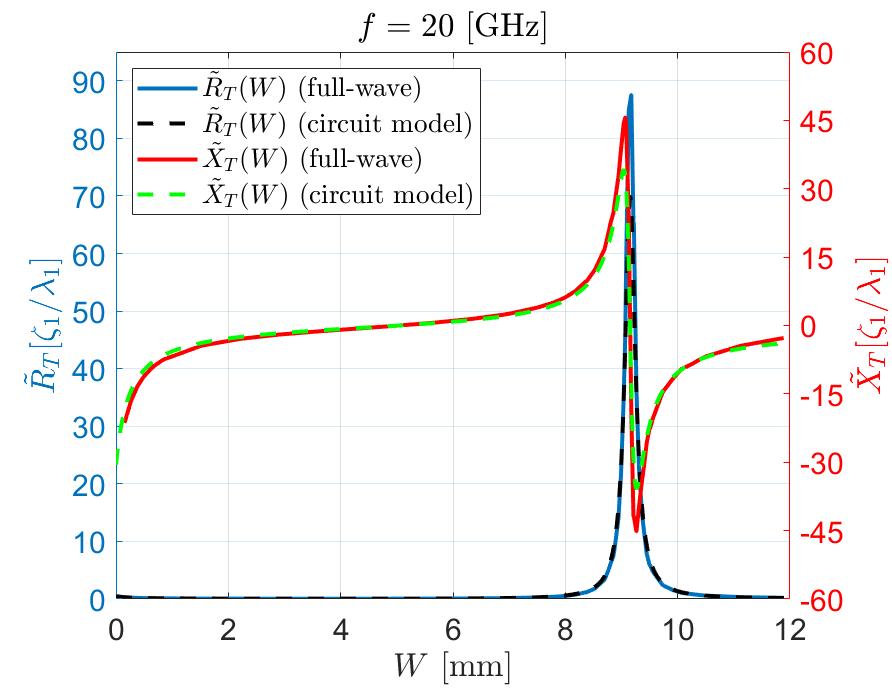}}
    \subfigure[]{\includegraphics[width=0.38\textwidth]{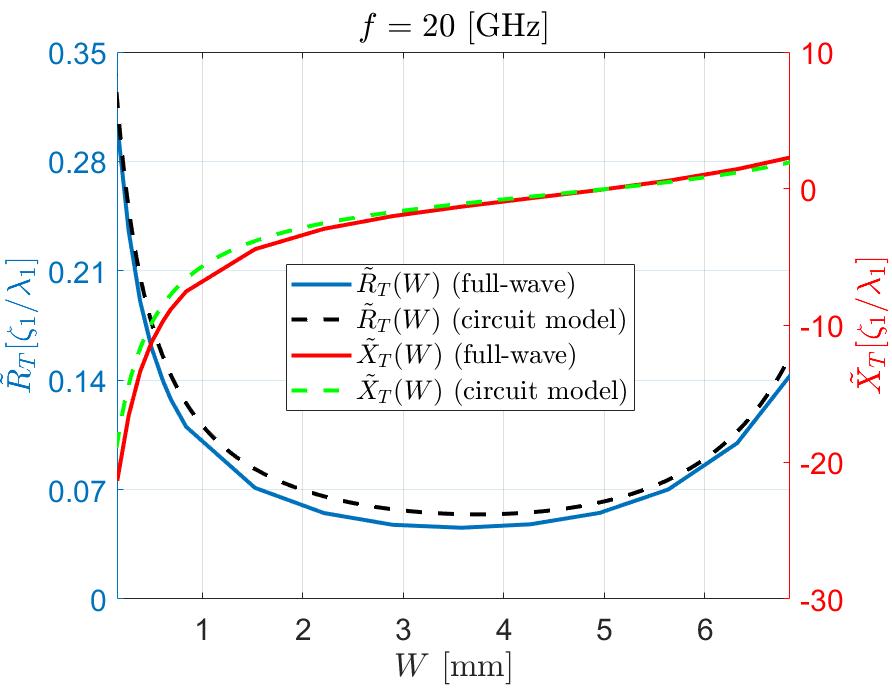}}
    \caption{\textcolor{black}{Validation of the} equivalent circuit \textcolor{black}{model of Fig. \ref{fig:EqvCirc} at the operating frequency $f=20$ GHz with the parameters evaluated in Section \ref{subsub:circuit_parameter_evaluation} (Table \ref{tab:CircuitParameters})}. (a) \textcolor{black}{Effective load resistance} $\tilde{R}_{T}(W)$ \textcolor{black}{(blue solid and black dashed lines) and reactance} $\tilde{X}_{T}(W)$ \textcolor{black}{(red solid and green dashed lines) as evaluated via full-wave simulations (procedure in \ref{subsec:dielectric_loss}, solid lines) and as estimated from the equivalent circuit (dashed lines) following \eqref{Eq:Imped} with the parameters Table \ref{tab:CircuitParameters}.} (b) Zoom in on \textcolor{black}{the} range \textcolor{black}{$W\in[0,6]\,\mathrm{mm}$.}} 
    \label{fig:LUT_fitting}
\end{figure}

To provide \textcolor{black}{quantitative} validation of the developed circuit model, we use the parameters extracted in Section \ref{subsub:circuit_parameter_evaluation} \textcolor{black}{(Table \ref{tab:CircuitParameters})} with \eqref{Eq:Imped} to predict the effective load impedance trends as a function of $W$ and compare them with full-wave simulation results (ANSYS HFSS). The results, presented in Figs. \ref{fig:LUT_fitting}(a) and (b), demonstrate \textcolor{black}{very good} correspondence between the equivalent circuit model and the data recorded in full-wave simulations, for both the resistive and reactive components. \textcolor{black}{This agreement, obtained} despite the very basic and intuitive capacitance and inductance models \textcolor{black}{\eqref{Eq:CircuitParameters}} used in \textcolor{black}{the previous subsection and the approximations involved in Steps \ref{itm:Rc}-\ref{itm:L0} therein}, implies that the equivalent circuit indeed captures the dominant physical phenomena governing the meta-atom response, as laid out in Section \ref{subsub:physical_considerations}.

\begin{figure*}
    \centering    
    \subfigure[]{\includegraphics[width=0.38\textwidth]{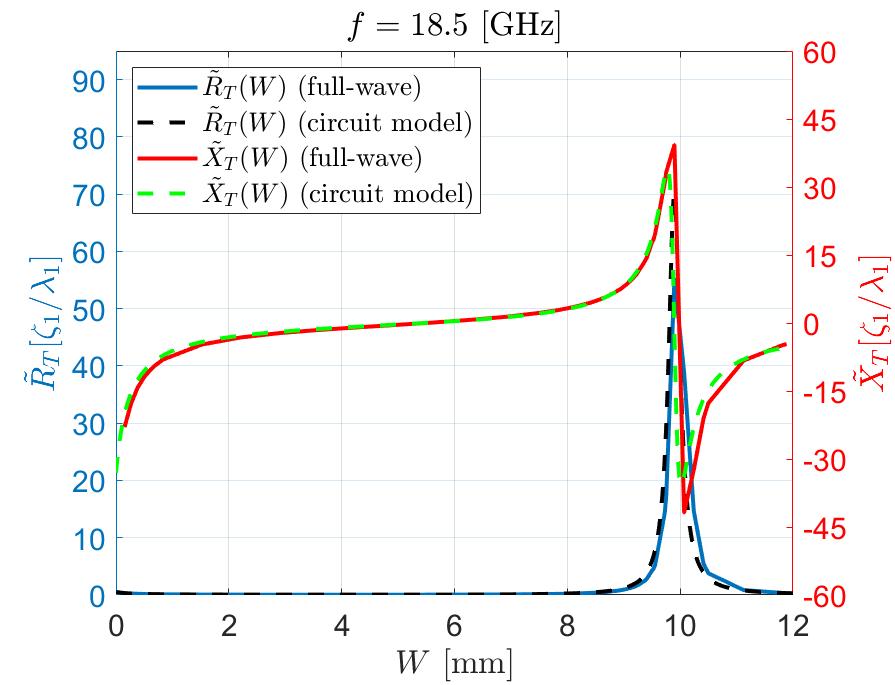}}
    \subfigure[]{\includegraphics[width=0.38\textwidth]{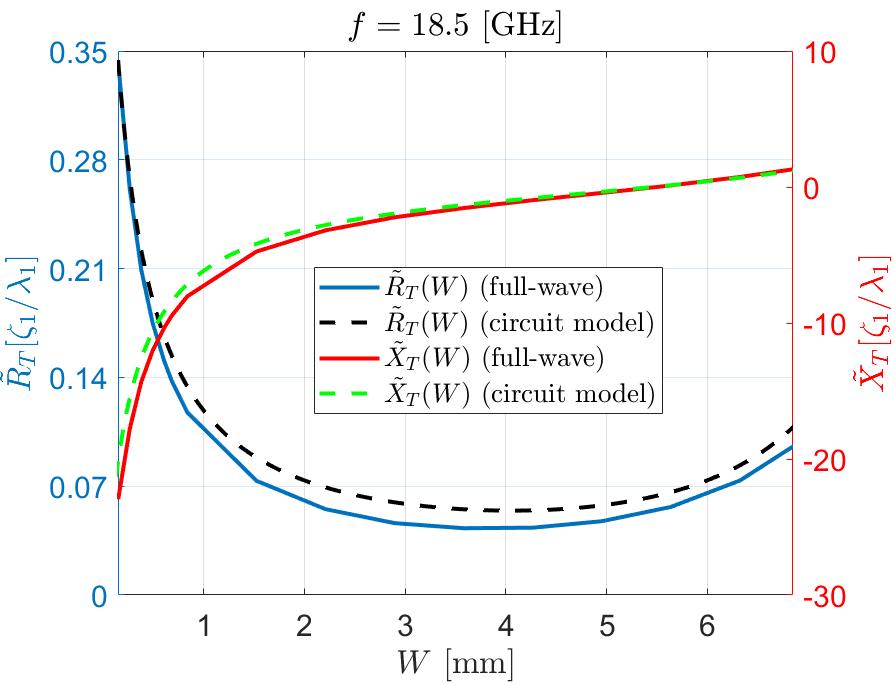}}    
    \subfigure[]{\includegraphics[width=0.38\textwidth]{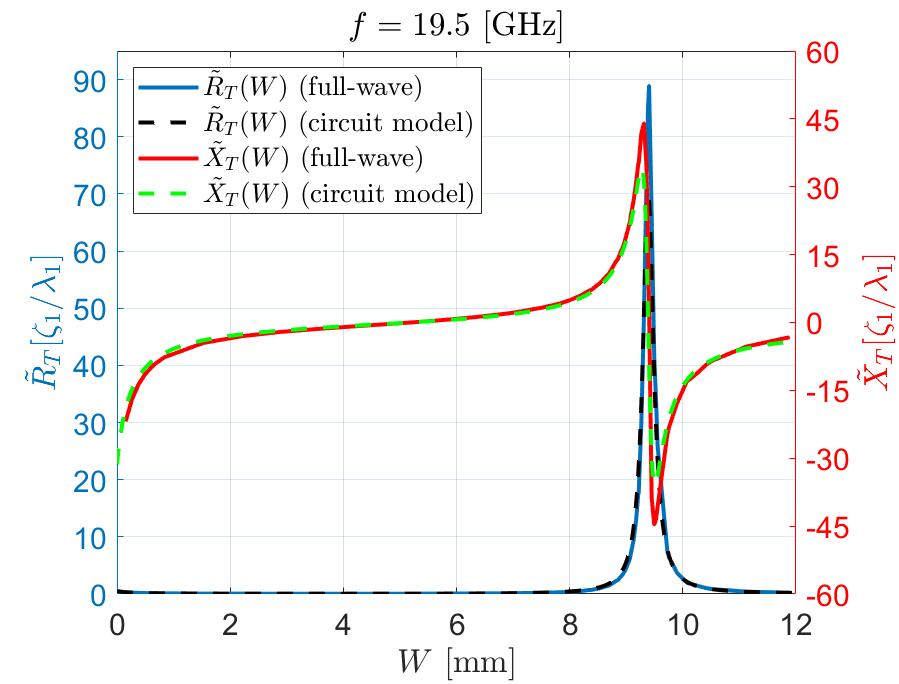}}
    \subfigure[]{\includegraphics[width=0.38\textwidth]{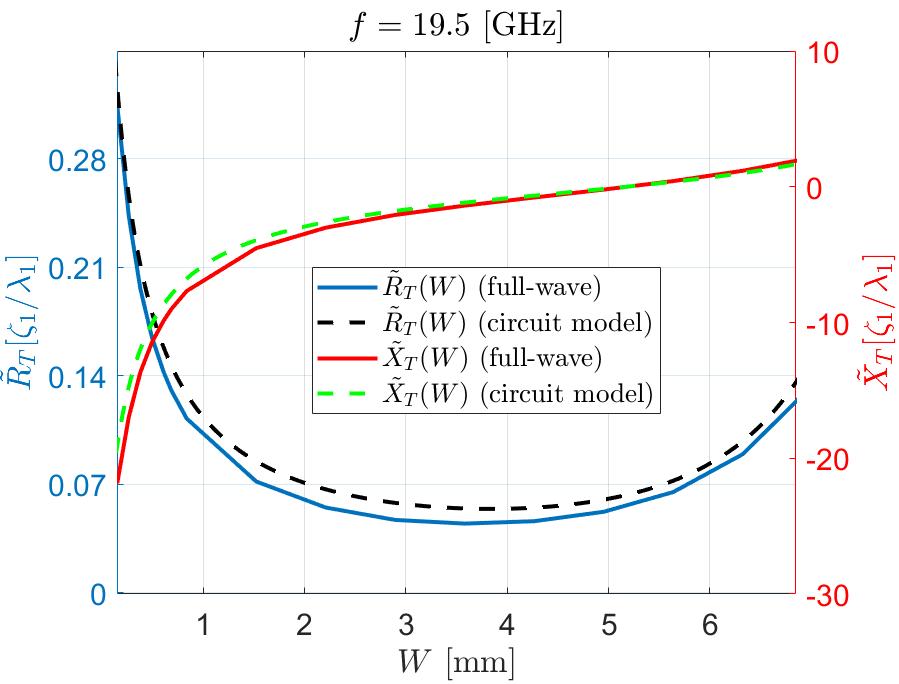}}    
    \subfigure[]{\includegraphics[width=0.38\textwidth]{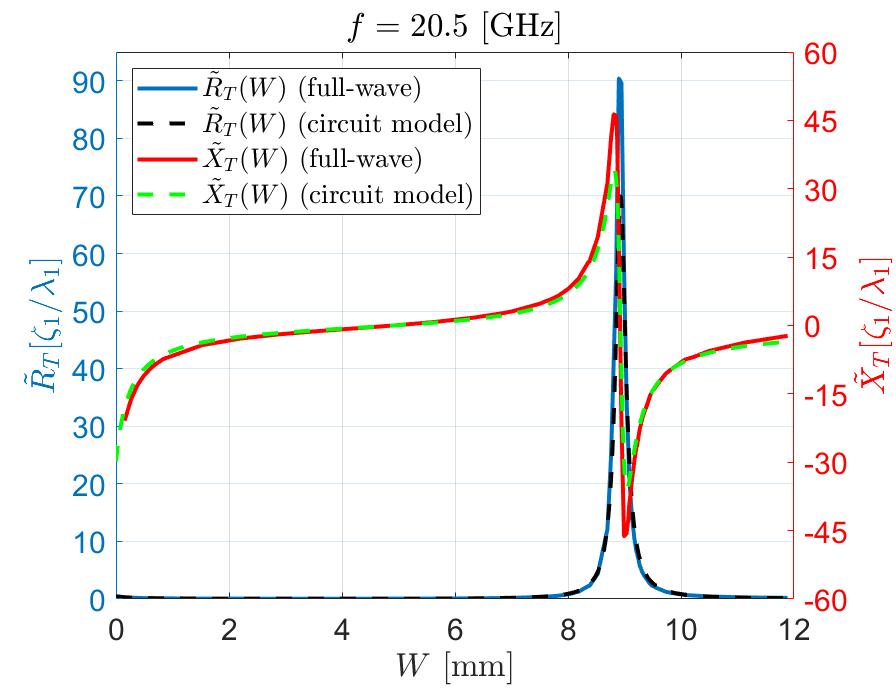}}    
    \subfigure[]{\includegraphics[width=0.38\textwidth]{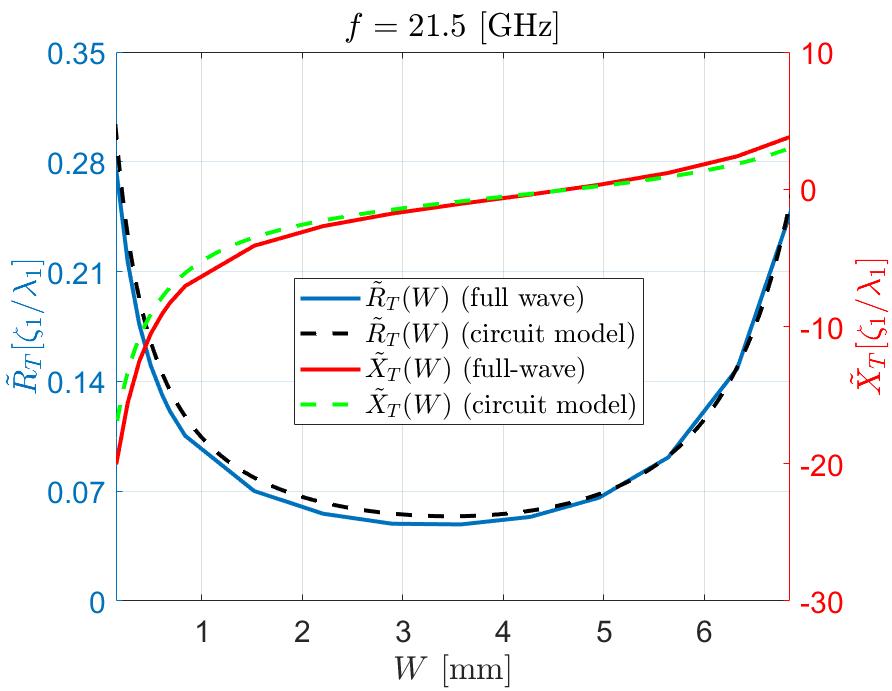}}
    \subfigure[]{\includegraphics[width=0.38\textwidth]{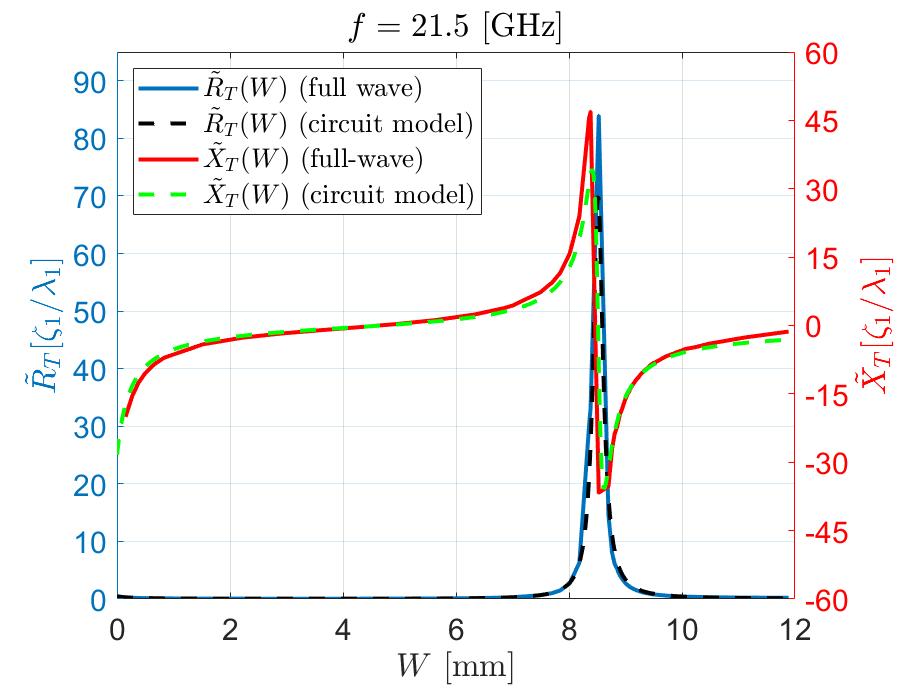}}
    \subfigure[]{\includegraphics[width=0.38\textwidth]{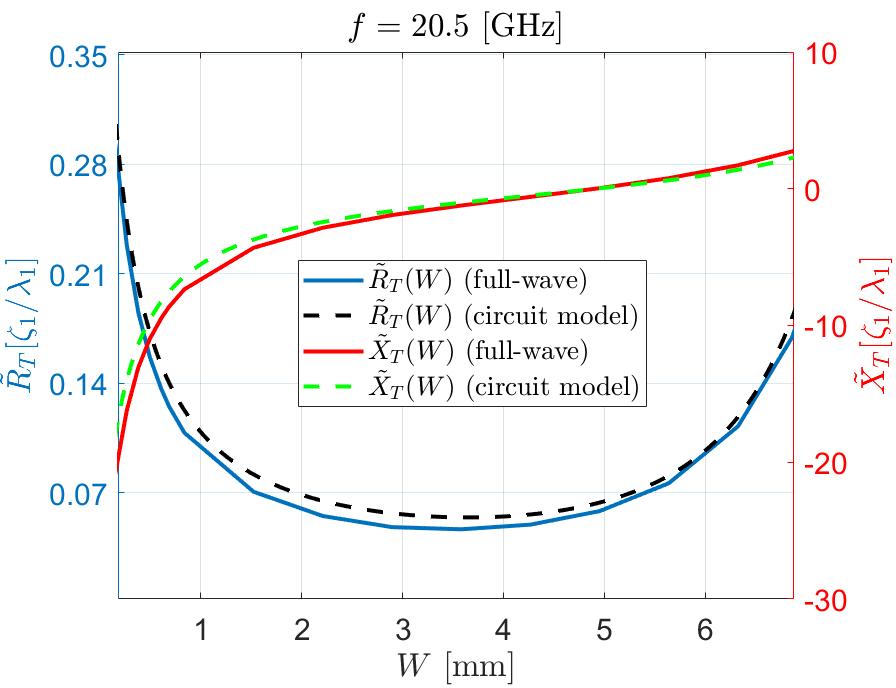}}
    \caption{\textcolor{black}{
    Effective load resistance $\tilde{R}_{T}(W)$ (blue solid and black dashed lines) and reactance $\tilde{X}_{T}(W)$ (red solid and green dashed lines) as evaluated via full-wave simulations (procedure in \ref{subsec:dielectric_loss}, solid lines) and as estimated from the equivalent circuit (dashed lines) following \eqref{Eq:Imped} with the parameters Table \ref{tab:CircuitParameters} for (a),(b) $f=18.5$ GHz; (c),(d) $f=19.5$ GHz; (e),(f) $f=20.5$ GHz; (g),(h) $f=21.5$ GHz. (a),(c),(e),(g) Entire considered capacitor widths $W\in[0,12]\,\mathrm{mm}$; (b),(d),(f),(h) Zoom in on the range $W\in[0,6]\,\mathrm{mm}$.}}
    \label{fig:Rt_Xt}
\end{figure*}

\begin{figure}
    \includegraphics[width=0.45\textwidth]{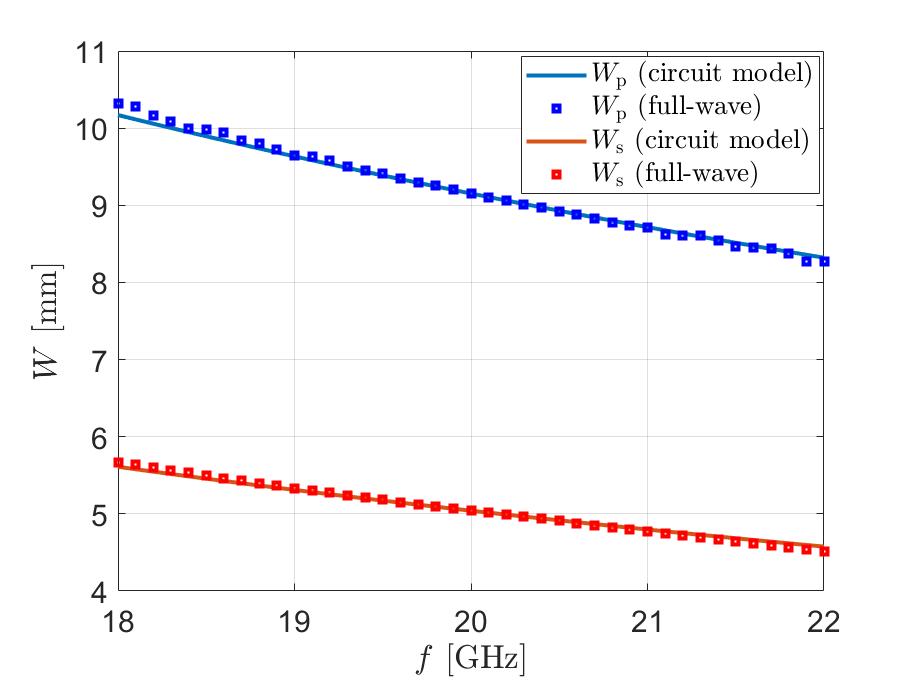}
    \centering
    \caption{\textcolor{black}{Parallel (blue) and series (red) RLC circuit (Fig. \ref{fig:EqvCirc}) resonant widths ($W_p$ and $W_s$, respectively) as a function of frequency, as extracted from wideband full-wave frequency simulations (square markers) and as predicted from the equivalent circuit model (solid lines) via  \eqref{Eq:Imped}. The resonant widths at a given frequency $f_0$ are defined as the zero crossing points of the total effective reactance $X_T(W;\omega=2\pi f_0)$.}}
    \label{fig:Resonances}
\end{figure}

To further establish the validity of the circuit model and the robustness of the identified underlying mechanisms, we probe the quality of its predictions at operating conditions outside the data range used to evaluate its parameters in Section  \ref{subsub:circuit_parameter_evaluation}. In particular, for the range of capacitor width $W\in[0,12]$ mm, we use the model of \eqref{Eq:Imped} with the exact same parameters as extracted at $f=20$ GHz  (\textcolor{black}{Table \ref{tab:CircuitParameters}}) to predict the effective distributed impedance $\tilde{Z}(W;\omega)$ in other operating frequencies. \textcolor{black}{Correspondingly, Fig. \ref{fig:Rt_Xt} presents the width dependency of total effective resistance} $\tilde{R}_T(W)$ \textcolor{black}{and reactance} $\tilde{X}_T(W)$ \textcolor{black}{at representative frequencies} within a fractional bandwidth of $20\%$ (between $18$ and $22$ GHz). \textcolor{black}{As can be seen, very} good agreement obtained between the circuit model and full-wave simulations \textcolor{black}{across the band:} the main resonant phenomena are clearly reproduced by the circuit, and impedance values (both resistive and reactive components) agree to a good extent with the numerical results. \textcolor{black}{These findings are further emphasized by the excellent correlation between the resonant widths of the equivalent series ($W_s$) and parallel ($W_p$) RLC circuit branches of Fig. \ref{fig:EqvCirc} as predicted by the circuit model and as identified from full-wave simulated data (Fig. \ref{fig:Resonances}). As expected, these widths drift towards higher (lower) values as the operating frequency decreases (increases) due to the change in the effective electrical length of the printed traces, maintaining very good agreement for the examined band of frequencies} (resonant \textcolor{black}{widths} prediction accuracies of the order of $\sim1\%$ and better).


These observations stress the significance of the developed model - not only for analysis, but also for synthesis: despite being devised based on simulated data at a \emph{single} frequency, it can be used for the design of MGs across an extended bandwidth, for a wide range of capacitor widths. This forms a reliable physically insightful semianalytical framework, properly incorporating dielectric and conductor loss factors, for designing MGs with prescribed frequency responses, as required for contemporary communication and imaging applications. Moreover, identifying the dominant processes provides insights regarding the different possible operation regimes and their impact on the device performance, pointing out regions of predominantly small or large resistances and qualitatively different frequency response profiles, highly useful when meeting the demands of specific applications - be it for increased scattering efficiency or enhanced absorption, narrowband sensors or broadband radar cross section reduction apparatuses. 


\subsection{Low-cost MG prototype}
\label{subsec:prototype_design}


Having established the relationship between the load impedance characteristics and the various loss mechanisms in the different $W$ regimes, we can proceed with the anomalous reflection MG design based on the analytical model developed in Section \ref{sec:theory}. We recall that our goal is to devise an MG (consisting of two meta-atoms per priod) that would funnel the maximum possible power carried by a $f=20$ GHz TE-polarized plane wave from $\theta_\mathrm{in}=10^\circ$ towards a given $\theta_\mathrm{out}$, countering power dissipation in the utilized low-cost (lossy) FR4 substrate (Fig. \ref{fig:configuration}). To this end, we explore the MG parameter space - including the substrate thickness $h$, interelement horizontal spacing $d$, and the capacitor widths $W_1$ and $W_2$ - to find optimal solutions for each of the considered $\theta_\mathrm{out}$. In particular, 
for each combination of $h$ and $d$ within the intervals $[0,0.2\lambda]$ and $[0.1\Lambda,0.9\Lambda]$, respectively, 
we calculate using \eqref{Eq:I0_I1}, \eqref{Eq:anomalEff} and $\tilde{Z}\textcolor{black}{_T}(W)$ of Fig. \ref{fig:ExtrConfig1andLUT}\textcolor{black}{(b) (corresponding to \eqref{Eq:Imped} at $f=20$ GHz)} the set of $W_\mathrm{1}$ and $W_\mathrm{2}$ [$\tilde{Z}_{1}=\tilde{Z}\textcolor{black}{_T}(W_1)$ and $\tilde{Z}_{2}=\tilde{Z}\textcolor{black}{_T}(W_2)$] that maximize anomalous efficiency $\eta_{-1}$. The configuration $(h,d,W_1,W_2)$ that yields the highest performance is designated as the optimal design (e.g., for prototype realization). 

It should be noted that even though the evaluation of the relation $\tilde{Z}\textcolor{black}{_T}(W)$ was performed across a broader range of $W$ in order to gain physical insight and identify the dominant mechanisms affecting this dependency (Fig. \ref{fig:LUT_fitting}\textcolor{black}{(a), Fig. \ref{fig:Rt_Xt}(a),(c),(e),(g),} and Section \ref{subsec:circuit_model}), we constrain ourselves to using only meta-atoms featuring small-to-moderate capacitor widths (below $W\approx 5$ mm $\approx \lambda/3$) in the ultimate prototype design. This is since the entire analytical modeling (Section \ref{sec:theory}) relies on the fact that the printed load features can be treated effectively as lumped impedances in a sense that the main interelement near-field coupling effects (mutual coupling between different loaded wires) stems from the dipolar term of the current distribution (the average current flowing along the $x$ axis); this manifests itself in the choice of Green's function for the analysis \cite{Rabinovich2018,Popov2018,8892735}. Nonetheless, when the printed capacitor \textcolor{black}{legs} grow larger, the current distributions may deviate substantially from this ansatz, giving rise to multipolar contributions that reduce the model accuracy (especially when the elements within the period become close to one another). Since the inherently inductive narrow copper strips naturally require capacitive loading in order to allow meaningful tuning of their response within a reasonable dynamic range, limiting thus the considered geometries to $W<5$ mm does not restrict the achievable performance of the design. In these cases where the semianalytical synthesis algorithm points out optimal load impedances that are close to the series equivalent circuit resonance ($X_T(W;f=20\mathrm{\,GHz})\approx0\Rightarrow W>5$ mm, \emph{cf.} Fig. \ref{fig:ExtrConfig1andLUT}\textcolor{black}{(b)}), we choose for the design a short-circuited strip of the nominal trace width (i.e., with $W=w=6$ mil and $s=0$), which well retain the dioplar approximation (due to the narrow geometry) while featuring the required near-resonance impedance\textcolor{black}{\footnote{\textcolor{black}{The minor capacitive value of the effective reactance extracted for this short-circuit (straight strip) geometry indicates that the flat wire approximation \cite{tretyakov2003analytical} used in our model to evaluate the self impedance of a current-carrying strip, utilizing an effective wire-equivalent radius of $r_\mathrm{eff}=w/4$ (e.g., in \eqref{Eq:I0_I1}), slightly overestimates the strip inductance.}}} 
$\tilde{Z}_{\mathrm{sc}}=(0.02 - j0.3) [\zeta_{1}/\lambda_{1}]$. 

\begin{figure*}[h!]
    \centering
    \subfigure[]{\includegraphics[width=0.32\textwidth]{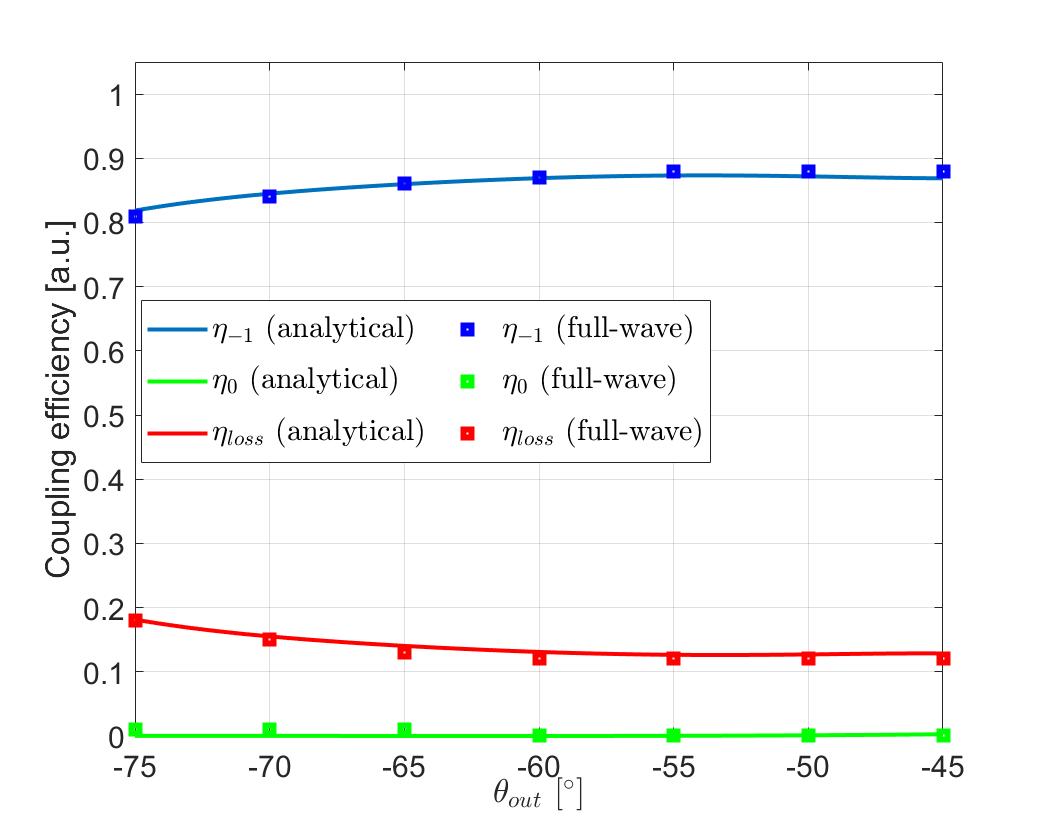}}
    \subfigure[]{\includegraphics[width=0.32\textwidth]{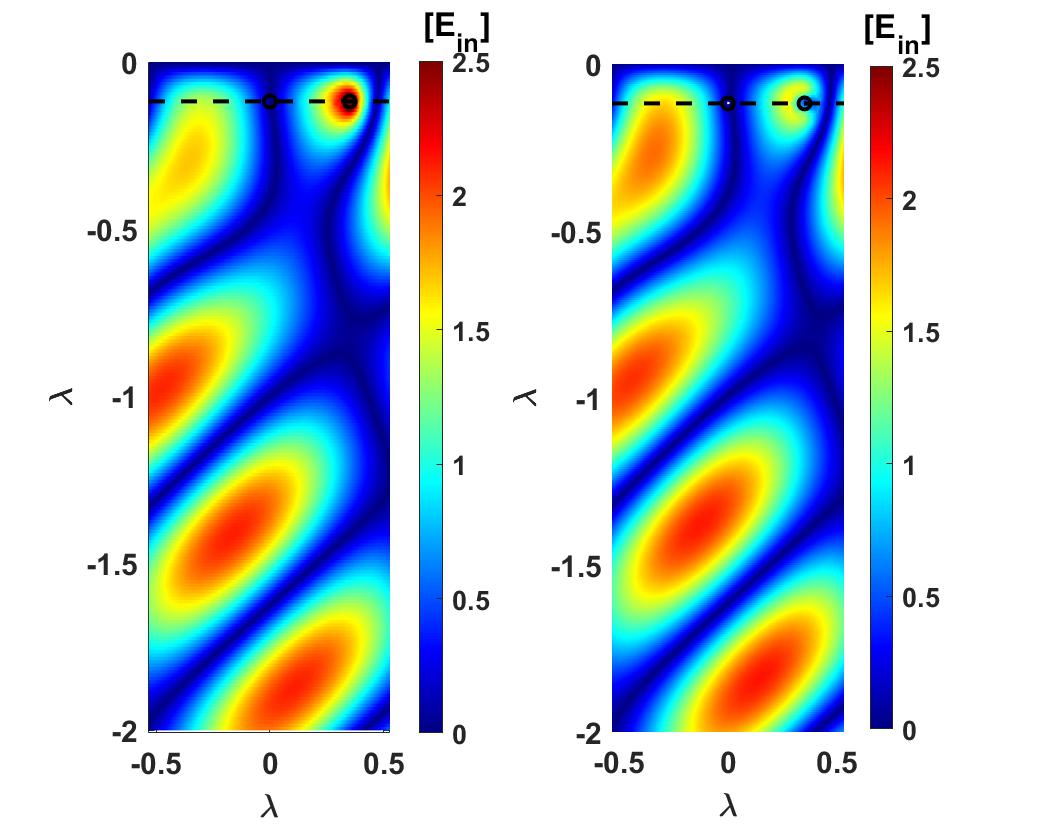}}
    \subfigure[]{\includegraphics[width=0.32\textwidth]{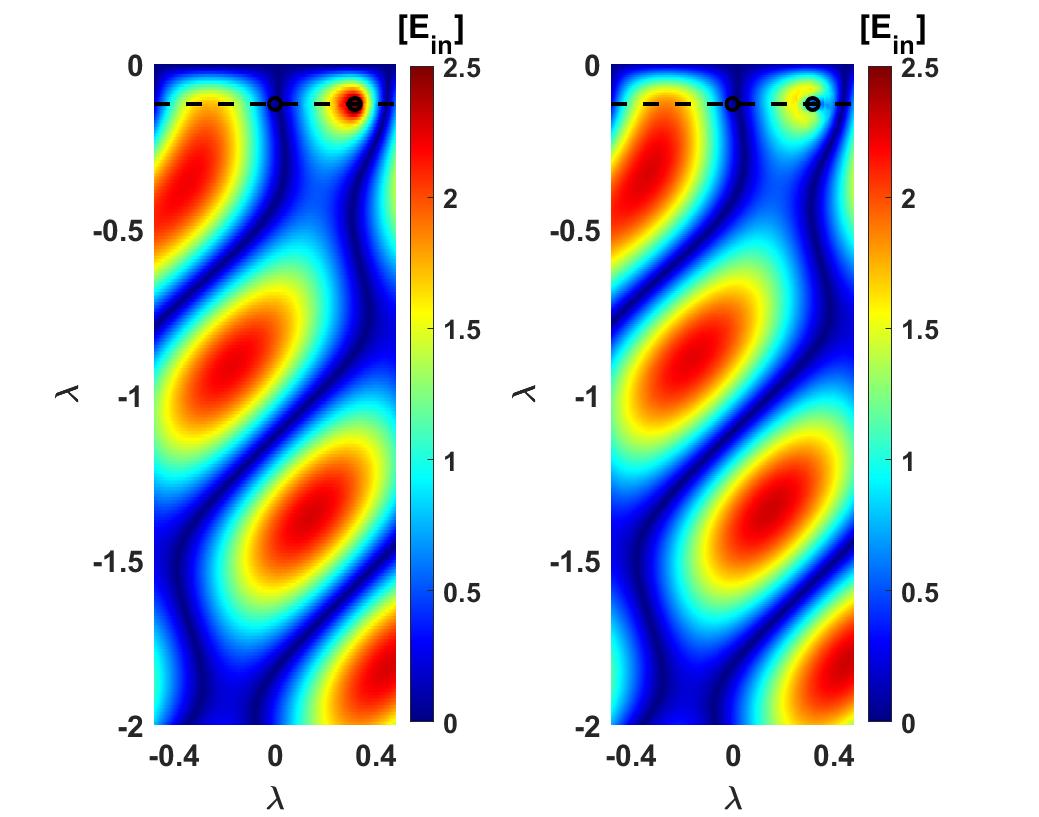}}
    \caption{Performance of the best-performing anomalous reflection low-cost PCB MG designs, deflecting power from $\theta_\mathrm{in}=10^\circ$ towards a prescribed $\theta_\mathrm{out}$\textcolor{black}{, corresponding to Table \ref{table:SimResults}}. (a) Analytical (solid lines) and full-wave (square markers) fraction of power absorbed (red) or coupled to the anomalous (blue) and specular (green) reflection modes is plotted as a function of the design reflection angle $\theta_\mathrm{out}$ (including effect of realistic FR4 and copper loss). \textcolor{black}{Total electric field snapshots $E_x(y,z)$ for the MGs designed for deflecting the incident wave towards (b) $\theta_\mathrm{out} = -50^\circ$ and (c) $\theta_\mathrm{out} = -60^\circ$,} comparing \textcolor{black}{the analytical predictions via \eqref{Eq:Tot_fields} and \eqref{Eq: SubsFields} (left) to the full-wave simulated results (right). A single period along $y$ is shown, with black horizontal dashed lines denoting the dielectric/air interface at $z=-h$, and black circles marking the locations of the loaded strips (meta-atoms).} 
    }
    \label{fig:results}
\end{figure*}

The optimal MG parameters obtained following this scheme, for \textcolor{black}{the prescribed angle of incidence $\theta_\mathrm{in}=10^\circ$ and} several values of deflection angles $\theta_\mathrm{out}$, are given in Table \ref{table:SimResults} along with relevant performance \textcolor{black}{figures} as evaluated from full-wave simulations for each of the proposed designs. These characteristics, quantifying the fraction of power coupled to the desired ($m=-1$, anomalous reflection) and undesired ($m=0$, specular reflection) modes, as well as the fraction of dissipated power (loss), are shown in Fig. \ref{fig:results}(a) as a function of $\theta_\mathrm{out}$. It can be observed that, despite the lossy substrate, the anomalous reflection efficiency $\eta_\mathrm{-1}$ remains 
greater than $80\%$ across the entire range of examined output angles, with efficiencies above $87\%$ for $|\theta_\mathrm{out}|\leq60^\circ$. Furthermore, there is a strong agreement between the results predicted by the analytical model and those as recorded by ANSYS HFSS with the semianalytically obtained MG design, verifying our theoretical approach and its treatment of dielectric loss (Sections \ref{sec:theory}, \ref{subsec:dielectric_loss}, and \ref{subsec:circuit_model}). 

\begin{table*}[t]
\centering
\caption{Design specifications and simulated performance of the designed \textcolor{black}{MGs operating} at $f=20 \mathrm{GHz}$ \textcolor{black}{(corresponding to Fig. \ref{fig:results})}.}
\begin{tabular}{ l m{1.5cm} m{1.5cm} m{1.5cm} m{1.5cm} m{1.5cm} m{1.5cm} m{1.5cm} } 
 \hline\hline
 $\theta_\mathrm{out}$ & $-75^\circ$ & $-70^\circ$ & $-65^\circ$ & $-60^\circ$ & $-55^\circ$ & $-50^\circ$ & $-45^\circ$\\ [0.5ex]
 \hline
 $\Lambda[\lambda]$ & 0.880 & 0.900 & 0.927 & 0.961 & 1.007 & 1.067 & 1.134\\
 
 $h[\lambda]$ & 0.126 & 0.124 & 0.122 & 0.121 & 0.119 & 0.118 & 0.117\\

 $d[\Lambda]$ & 0.30 & 0.31 & 0.32 & 0.33 & 0.33 & 0.33 & 0.33\\
 
 $W_1$[mm] & $0.152\textcolor{black}{^\dagger}$ & $0.152\textcolor{black}{^\dagger}$ & $0.152\textcolor{black}{^\dagger}$ & $0.152\textcolor{black}{^\dagger}$ & $0.152\textcolor{black}{^\dagger}$ & $0.152\textcolor{black}{^\dagger}$ & $0.152\textcolor{black}{^\dagger}$\\
 
 $W_2$[mm] & 0.997 & 1.064 & 1.131 & 1.172 & 1.212 & 1.2253 & 1.2387\\
 
 Anomalous reflection ($\eta_{-1}$) & $81.7\%$ & $84.5\%$ & $86.0\%$ & $86.9\%$ & $87.3\%$ & $87.2\%$ & $86.9\%$\\
 
 Specular reflection ($\eta_{0}$) & $0.0\%$ & $0.0\%$ & $0.0\%$ & $0.0\%$ & $0.0\%$ & $0.1\%$ & $0.3\%$\\
 
 Losses & $18.3\%$ & $15.5\%$ & $14.0\%$ & $13.1\%$ & $12.7\%$ & $12.7\%$ & $12.8\%$\\ 
 \hline\hline
\end{tabular}

\raggedright \textcolor{black}{$^\dagger$ As discussed in the beginning of Section \ref{subsec:prototype_design}, when the semianalytical synthesis algorithm points out optimal load impedances that are close to the series equivalent circuit resonance ($X_T(W;f=20\mathrm{\,GHz})\approx0\Rightarrow W>5$ mm), a short-circuited strip of the nominal trace width (i.e., with $W=w=6\,\mathrm{mil}=0.152\,\mathrm{mm}$ and $s=0$) is chosen for the design.}
\label{table:SimResults}
\end{table*}

Further evidence to this agreement can be seen in Fig. \ref{fig:results}(b) and (c), presenting snapshots of the spatial field distributions 
\textcolor{black}{for} two representative configurations, corresponding to the designs for anomalous reflection towards $\theta_\mathrm{out}=-50^\circ$ and $\theta_\mathrm{out}=-60^\circ$, respectively (Table \ref{table:SimResults}). Once again, excellent correspondence between the analytical predictions and full-wave simulations is obtained, except for the regions surrounding the meta-atoms\footnote{Since the model does not take into account the finite dimensions of the printed capacitors, some discrepancy around these regions is expected \cite{epstein2017unveiling}.}. 

\begin{figure}[t!]
    \includegraphics[width=0.38\textwidth]{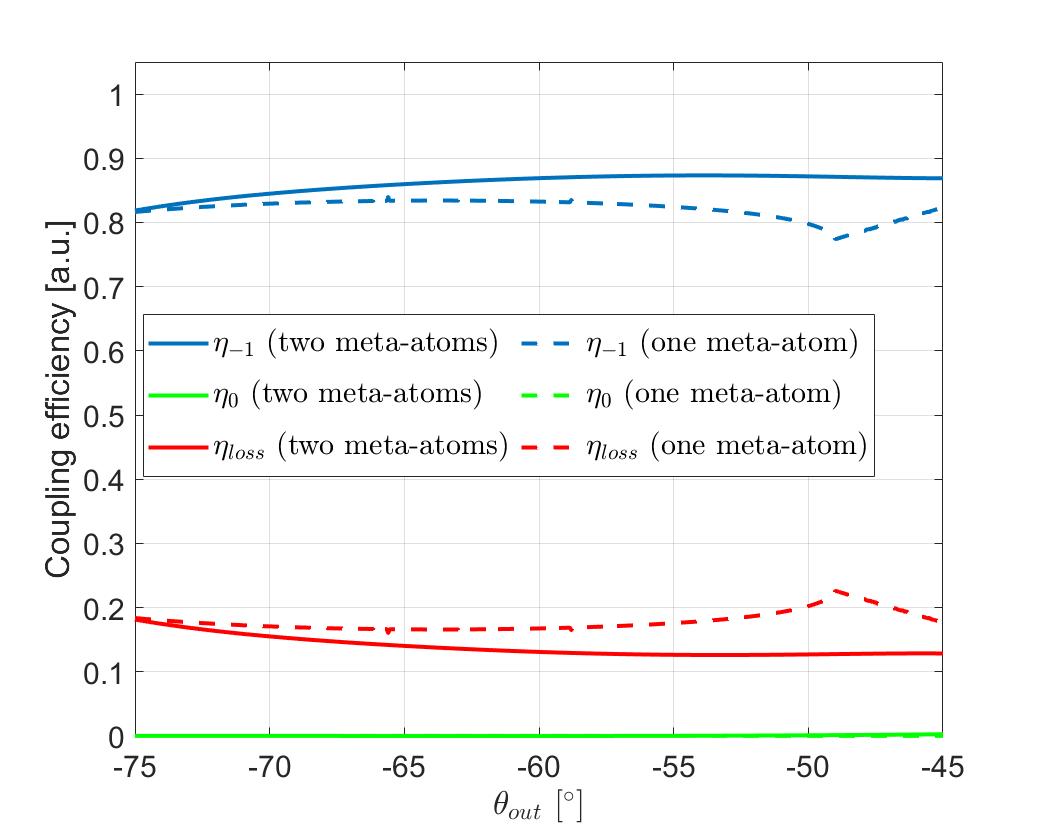}
    \centering
    \caption{\textcolor{black}{Analytically predicted fraction of power coupled to anomalous reflection ($\eta_{-1}$ of \eqref{Eq:anomalEff}, blue), specular reflection ($\eta_0$, \eqref{Eq:specEff}, green), and absorbed by conductors and dielectrics ($\eta_\mathrm{loss}$, \eqref{Eq:loss}, red) of the optimal MG designs obtained via the methodology presented in Section \ref{sec:theory} with the configuration in Fig. \ref{fig:configuration} having either one (dashed lines) or two (solid lines) meta-atoms per period.}} 
    \label{fig:comparPerf}
\end{figure}

\textcolor{black}{Before we conclude our discussion in the prototype design, it would be worthwhile to review our choice (Fig. \ref{fig:configuration} and Section \ref{sec:theory}) of using two meta-atoms per period for the low-cost MGs under consideration. Indeed, as mentioned in Section \ref{subsec:scattered_fields}, for ideally lossless configurations, it has been shown in \cite{Rabinovich2018} that a single meta-atom per period is sufficient to obtain perfect anomalous reflection for the scenarios considered herein. Nonetheless, when attempting to apply this naive approach with a lossy low-cost substrate such as FR4 (i.e., maximizing $\eta_{-1}$ of \eqref{Eq:anomalEff} with $I_2=0$, leaving only $h$ and $W_1$ as degrees of freedom), we obtained the results described by the dashed lines in Fig. \ref{fig:comparPerf}. As can be seen, although spurious (specular) reflection is kept at a minimum for all designs, deflecting the beam incoming from $\theta_\mathrm{in}=10^\circ$ to the various $\theta_\mathrm{out}$, losses are considerably high, limiting anomalous reflection efficiencies. Since increasing the number of available degrees of freedom has been utilized in the past to overcome conductor loss in MG synthesis (e.g., see \cite{8892735}), we decided to try this approach herein as well as a means to mitigate dielectric loss. As verified by the solid lines in Fig. \ref{fig:comparPerf} (identical to those in Fig. \ref{fig:results}(a)), the introduction of a second meta-atom per period (offering $h$, $d$, $W_1$, and $W_2$ of Fig. \ref{fig:configuration} for the design) indeed improved the achievable anomalous reflection efficiencies on the expense of power dissipation. Since the added loaded strip does not complicate the fabrication (the MG still forms a low-cost single-layer FR4 PCB), we chose this superior bipartite configuration as the basis for our MG synthesis procedure in this work.}

Overall, these results indicate the high fidelity of the developed model, augmented as to properly account for all dissipation factors in the system. Importantly, it points out that the inclusion of sufficient degrees of freedom and suitable load modeling indeed allows overcoming nontrivial dielectric loss in low-cost substrate and achieve \textcolor{black}{considerably} high diffraction efficiencies in loaded-wire MGs. By harnessing the evanescent waves inherently including in the detailed formulation as means to redistribute the power within the period as to minimize the effective currents induced on each of the meta-atoms, while choosing a substrate thickness that avoids resonant multiple reflections within the lossy dielectric (as also highlighted in \cite{diker2023low}), 
one may realize effective beam manipulation MGs with substantially reduced cost and complexity.

\subsection{Experiment}
\label{subsec:experiment}
\begin{figure}[t!]
    \includegraphics[width=0.45\textwidth]{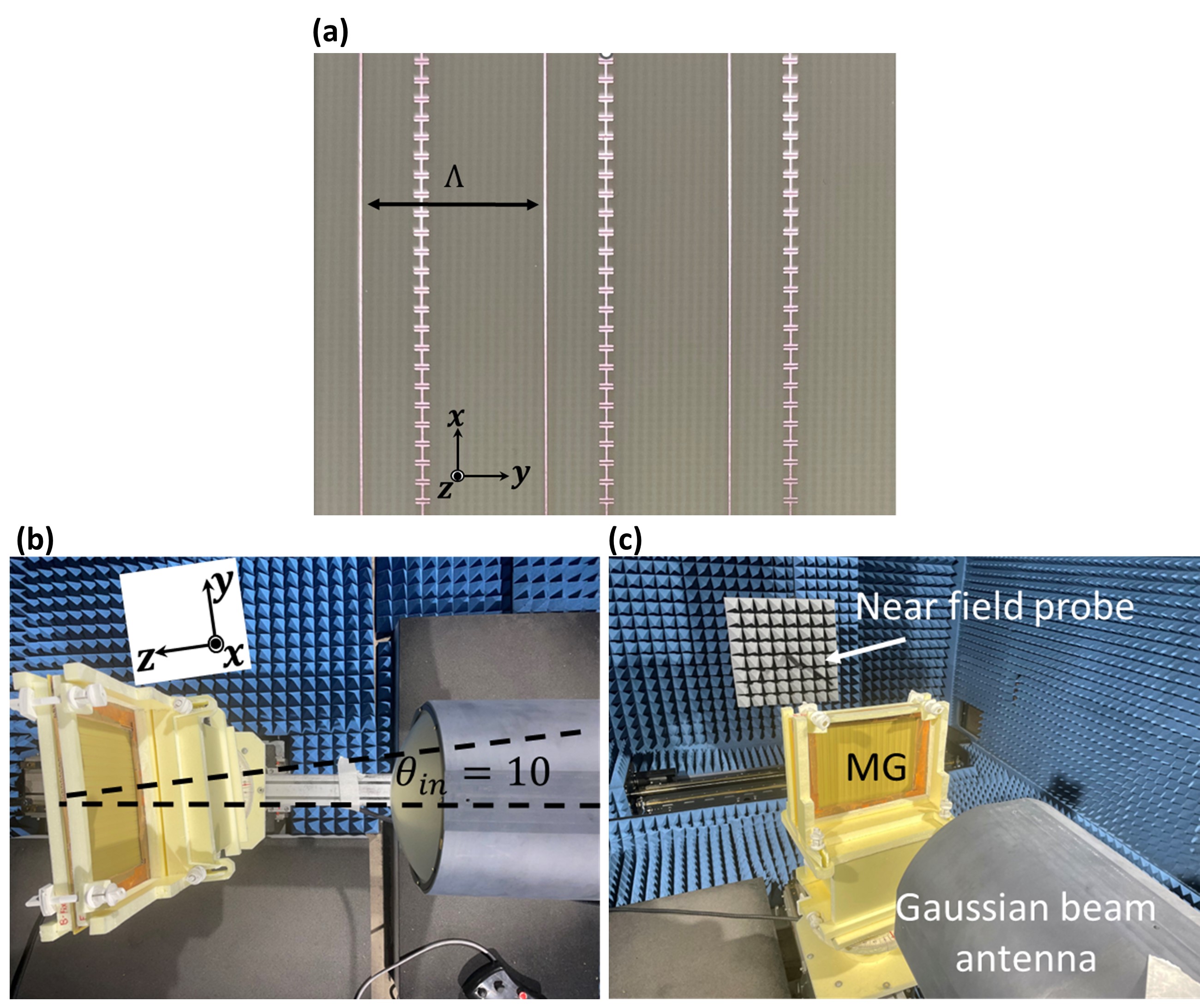}
    \centering
    \caption{\textcolor{black}{E}xperimental verification \textcolor{black}{of the prototype manufactured according to the design in Section \ref{subsec:prototype_design}, corresponding to Fig. \ref{fig:results}(b)}. (a) \textcolor{black}{Top view of} the fabricated MG \textcolor{black}{(}zoom in on \textcolor{black}{a section of area $24l\times3\Lambda$). (b) Top view and (c) side view of the experimental setup in the anechoic chamber, including a} Gaussian beam antenna, near-field measurement system\textcolor{black}{,} and \textcolor{black}{the MG (}DUT\textcolor{black}{)}. 
    }
    \label{fig:Setup}
\end{figure}

Upon validating the model's accuracy, we proceed towards experimental verification of the proposed approach. In particular, we consider the design parameters of Table \ref{table:SimResults} associated with the MG aimed at redirecting a plane wave incoming from $\theta_\mathrm{in}=10^\circ$ towards $\theta_\mathrm{out}=-50^\circ$ in reflection [Fig. \ref{fig:results}(b)]. Correspondingly, we have fabricated a 
$9''\times12''$ MG prototype using standard PCB technology, shown in Fig. \ref{fig:Setup}(a). 

\begin{figure}[t!]
    \includegraphics[width=0.4\textwidth]{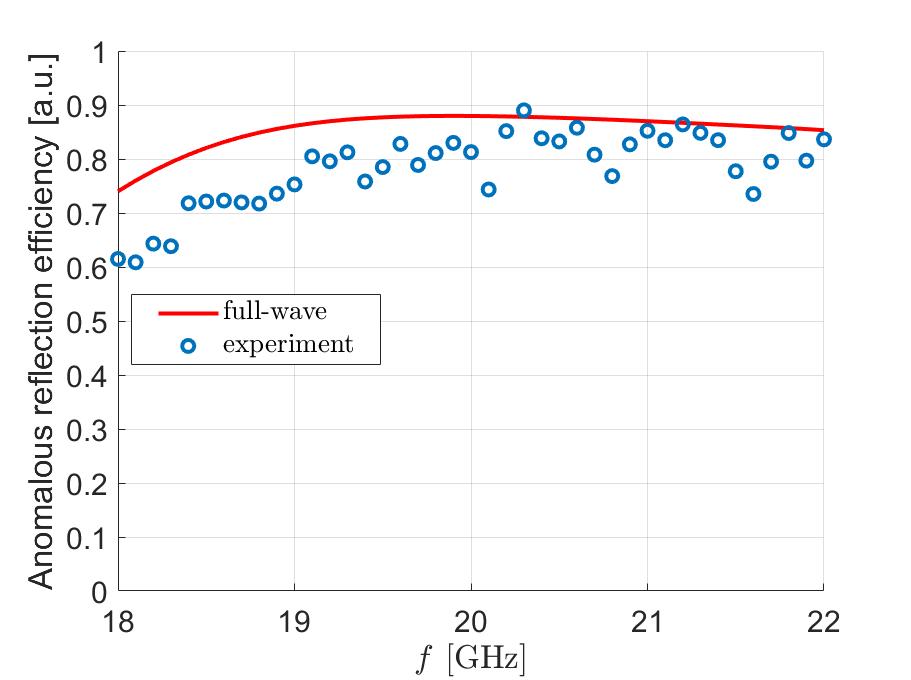}
    \centering
    \caption{\textcolor{black}{Fraction of incident power coupled to the anomalous reflection mode $\eta_{-1}$} as a function of frequency\textcolor{black}{, as measured experimentally for the fabricated prototype of Fig. \ref{fig:Setup}} (blue circle \textcolor{black}{markers}) \textcolor{black}{and as recorded in} full-wave simulation (solid red line)\textcolor{black}{.}}
    \label{fig:AnomExp}
\end{figure}

The MG was characterized in the anechoic chamber at the Technion. The device under test (DUT) was positioned \textcolor{black}{on a foam holder in front} of a Gaussian beam antenna (Millitech, Inc., GOA-42-S000094, focal distance of $196 \mathrm{\,mm} \approx 13\lambda$) and illuminated by a quasi-planar wavefront at $\theta_\mathrm{in}=10^\circ$ \textcolor{black}{[}Fig. \ref{fig:Setup}(b)]. A cylindrical near-field measurement system (MVG/Orbit-FR) was employed to quantify the scattered field \textcolor{black}{at a constant radius} around the DUT. \textcolor{black}{This was achieved} by rotating the DUT and the antenna together (to keep the incident angle fixed) in front of \textcolor{black}{the} near field probe, \textcolor{black}{with the} distance \textcolor{black}{between the DUT and the probe remaining fixed at} $850 \mathrm{\,mm} \approx 57\lambda$ \textcolor{black}{[}Fig. \ref{fig:Setup}(c)]. \textcolor{black}{Due to blockage effects presented by the Gaussian beam antenna when scanning at wide angles, the MG was eventually placed at a larger distance than the one defined by the focal plane ($430\,\mathrm{mm}\approx29\lambda$). This is supported by observations from previous work, implying that the Gaussian beam illumination at this distance is still sufficiently collimated as to facilitate proper characterization of the device \cite{rabinovich2020dual}.} Once the near-field distribution has been obtained, the far-field radiation pattern was computed by the system based on the equivalence principle \cite{balanis2012advanced}. 
For reference, the DUT was removed to measure the direct radiation of the source, enabling assessment of the excitation power\textcolor{black}{\footnote{\textcolor{black}{For both the reference and DUT measurements, a thin metallic frame matching the outer circumference of the prototype MG was mounted on the foam holder to guarantee that all reference fields measured in transmission indeed fully interacted with the core of the DUT and can be used for reliable calibration of the scattering pattern for efficiency calculations (Fig. \ref{fig:AnomExp}).}}}. 

\begin{figure}[t!]
    \includegraphics[width=0.4\textwidth]{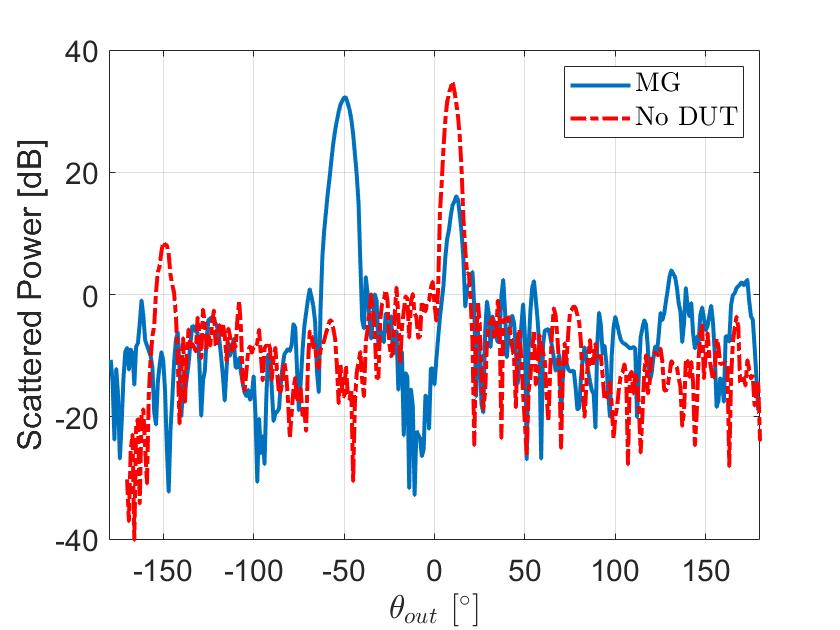}
    \centering
    \caption{\textcolor{black}{Measured scattering pattern for the prototype MG of Fig. \ref{fig:Setup}(a) and Fig. \ref{fig:results}(b), presenting the power reflected off the DUT when illuminated by a TE-polarized wave from $\theta_\mathrm{in}=10^\circ$ (Fig. \ref{fig:Setup}(b))} at the \textcolor{black}{designated} operating frequency of $f=20\textcolor{black}{\,\mathrm{GHz}}$ \textcolor{black}{(blue solid line). The reference pattern, recording the power received at the probe in the absence of the MG, is shown in dash-dotted red line.}} 
    \label{fig:Pattern}
\end{figure}

Figure \ref{fig:Pattern} 
presents the measured scattering pattern of the MG at the designated operation frequency $f=20$ GHz, along with the reference (source illumination) radiation pattern. As expected, when the MG (DUT) is absent (dashed red line), the plane-wave direct ray peak at $\theta_\mathrm{in}=10^\circ$ is clearly visible; on the other hand, when the incident plane wave from $\theta_\mathrm{in}$ impinges upon the MG (solid blue line), specular reflection ($m=0$ FB mode) is substantially suppressed, and the power is effectively funnelled towards the designated anomalous reflection ($m=-1$ FB mode) at $\theta_\mathrm{out}=-50^\circ$. 

Comparing in Fig. \ref{fig:AnomExp} the measured and simulated 
anomalous reflection efficiency $\eta_{-1}$ as a function of frequency\footnote{As in \cite{8892735}, in order to compare the scattering off the finite area experimental prototype and the full-wave simulations of the infinite periodic MG, we must factor out the differences in effective aperture size seen by the incident and reflected waves. Correspondingly, the experimental anomalous reflection efficiency is calculated from the measured data following $\eta_{-1,\mathrm{exp}} =  \frac{G_\mathrm{MG}(\theta_\mathrm{out})}{G_\mathrm{ref}(\theta_\mathrm{in})} \frac{\cos\theta_\mathrm{in}}{\cos\theta_\mathrm{out}}$, where $G_\mathrm{MG}(\theta_\mathrm{out})$ represents the gain measured from the scattered pattern of the DUT, while $G_\mathrm{ref}(\theta_\mathrm{in})$ denotes the gain measured from the scattered pattern of the reference \cite{8892735}.}, 
  reveals good agreement between theory (solid red line) and experiment (blue circle markers) within the band $[18,22]\mathrm{\,GHz}$. The minor discrepancies between the two curves may be attributed by the uncertainty in substrate parameters (low-cost FR4 composites often exhibit larger tolerances in both permittivity and loss tangent than their high-end laminate counterparts), as well as from the limited accuracy of the DUT manual alignment in terms of angle of incidence \cite{elad2024simultaneous}.
%
%
Quantitatively, the results indicate that anomalous reflection efficiencies of $81\%$ are obtained experimentally at a frequency of $20 \mathrm{\,GHz}$, retaining a rather stable and efficient beam deflection across a moderate bandwidth. 
These combined theoretical and experimental findings suggest that even in the presence of significant substrate loss, it is feasible - with the MG flexible degrees of freedom and analytical models - to identify working points that lead to high engineered diffraction efficiencies. 

\section{Conclusion}
\textcolor{black}{To conclude, we have presented a comprehensive modeling and implementation methodology for analysis and synthesis of loaded-wire MGs relying on lossy substrates, facilitating realization of low-cost PCB MGs for efficient anomalous reflection. Besides the practical implications of the proposed approach, enabling effective beam deflection through proper consideration of loss factors in the analytical derivation and the introduction of additional degrees of freedom to mitigate them, we proposed an equivalent circuit model to provide physical insight into the dominant power dissipation mechanisms and their effects on the device performance. Through an intuitive step-by-step analysis, it was shown that the behaviour of the load impedance as a function of the printed capacitor width can be well modelled by a combination of series and parallel resonant RLC circuits, tied to the fine physical features of the meta-atom geometry. Importantly, the devised detailed equivalent circuit points out that the non-negligible dielectric loss of the low-cost substrate manifests itself as effective (series and parallel) geometry-dependent resistive components - dominant over the constant one associated with conductor loss, exclusively considered in conventional MGs - which must be properly accounted for in order to reach an optimal operating point. Remarkably, the various equivalent circuit coefficients extracted from \emph{single-frequency} full-wave simulations using \emph{simplistic} component models to explain the load characteristics as a function of the printed capacitor width, were found to yield reliable quantitative predictions of the meta-atom resonances and complex effective impedance across a considerable band of frequencies as well - thus forming a highly useful semianalytical tool for wideband MG synthesis.}

\textcolor{black}{We harnessed these developments to design and demonstrate experimentally a simple, single-layer, low-cost FR4 PCB MG for anomalous reflection, achieving $>80\%$ power efficiency upon deflecting a} beam from $\theta_\mathrm{in}=10^\circ$ to $\theta_\mathrm{out}=-50^\circ$ \textcolor{black}{at $f=20$ GHz. The theoretical predictions based on the augmented analytical model, the full-wave numerical simulations, and the measured results from our near-field system, all match very well, further emphasizing the high fidelity of the developed seimanalytical synthesis scheme. These findings provide invaluable insights regarding the various effects of substrate and conductor loss on MG operation and analytical modeling, while paving the path for cost-effective realizations of this versatile complex media platform for efficient field manipulation. Beyond promoting integration in real-world microwave and antenna systems intended for mass production, the presented physical interpretation of the various power dissipation processes may find use in the development of enhanced PCB MG absorbers as well.}

\bibliographystyle{IEEEtran}
\bibliography{shorttitles,newBib}

\end{document}